\newcommand\apjcls{1}
\newcommand\aastexcls{2}
\newcommand\othercls{3}

\newcommand\papercls{\aastexcls}
\documentclass[tighten, times, twocolumn]{aastex62}  

\if\papercls \apjcls
\usepackage{apjfonts}
\else\if\papercls \othercls
\usepackage{epsfig}
\usepackage{margin}
\usepackage{times}
\fi\fi
\usepackage{ifthen}
\usepackage{natbib}
\usepackage{amssymb, amsmath}
\usepackage{appendix}
\usepackage{etoolbox}
\usepackage[T1]{fontenc}
\usepackage{paralist}

\if\papercls \apjcls
\newcommand\aas{\ref@jnl{AAS Meeting Abstracts}}
\newcommand\dps{\ref@jnl{AAS/DPS Meeting Abstracts}}
\newcommand\maps{\ref@jnl{MAPS}}
\else\if\papercls \othercls
\usepackage{astjnlabbrev-jh}
\fi\fi

\if\papercls \aastexcls
\hypersetup{citecolor=blue, 
            linkcolor=blue, 
            menucolor=blue, 
            urlcolor=blue}  
\else
\usepackage[
bookmarks=true,           
bookmarksnumbered=true,   
colorlinks=true,          
citecolor=blue,           
linkcolor=blue,           
menucolor=blue,           
urlcolor=blue,            
linkbordercolor={0 0 1},  
pdfborder={0 0 1},
frenchlinks=true]{hyperref}
\fi

\if\papercls \othercls

\else

\fi

\providecommand{\adsurl}[1]{\href{#1}{ADS}}

\makeatletter
\patchcmd{\NAT@citex}
  {\@citea\NAT@hyper@{%
     \NAT@nmfmt{\NAT@nm}%
     \hyper@natlinkbreak{\NAT@aysep\NAT@spacechar}{\@citeb\@extra@b@citeb}%
     \NAT@date}}
  {\@citea\NAT@nmfmt{\NAT@nm}%
   \NAT@aysep\NAT@spacechar\NAT@hyper@{\NAT@date}}{}{}

\patchcmd{\NAT@citex}
  {\@citea\NAT@hyper@{%
     \NAT@nmfmt{\NAT@nm}%
     \hyper@natlinkbreak{\NAT@spacechar\NAT@@open\if*#1*\else#1\NAT@spacechar\fi}%
       {\@citeb\@extra@b@citeb}%
     \NAT@date}}
  {\@citea\NAT@nmfmt{\NAT@nm}%
   \NAT@spacechar\NAT@@open\if*#1*\else#1\NAT@spacechar\fi\NAT@hyper@{\NAT@date}}
  {}{}
\makeatother

\makeatletter
\DeclareRobustCommand{\lowcase}[1]{\@lowcase#1\@nil}
\def\@lowcase#1\@nil{\if\relax#1\relax\else\MakeLowercase{#1}\fi}
\makeatother

\DeclareSymbolFont{UPM}{U}{eur}{m}{n}
\DeclareMathSymbol{\umu}{0}{UPM}{"16}
\let\oldumu=\umu
\renewcommand\umu{\ifmmode\oldumu\else\math{\oldumu}\fi}

\if\papercls \othercls

\else

\fi

\let\oldsim=\sim
\renewcommand\sim{\ifmmode\oldsim\else\math{\oldsim}\fi}
\let\oldpm=\pm
\renewcommand\pm{\ifmmode\oldpm\else\math{\oldpm}\fi}
\newcommand\by{\ifmmode\times\else\math{\times}\fi}

\newbox{\wdbox}
\renewcommand\c{\setbox\wdbox=\hbox{,}\hspace{\wd\wdbox}}
\renewcommand\i{\setbox\wdbox=\hbox{i}\hspace{\wd\wdbox}}

\newcount\timect
\newcount\hourct
\newcount\minct
\newcommand\now{\timect=\time \divide\timect by 60
         \hourct=\timect \multiply\hourct by 60
         \minct=\time \advance\minct by -\hourct
         \number\timect:\ifnum \minct < 10 0\fi\number\minct}

\catcode`@=11

\newcommand\comment[1]{}

\newcommand\commenton{\catcode`\%=14}

\renewcommand\math[1]{$#1$}
\newcommand\mathshifton{\catcode`\$=3}

\let\atab=&
\newcommand\atabon{\catcode`\&=4}

\let\oldmsp=\sp
\let\oldmsb=\sb
\def\sp#1{\ifmmode
           \oldmsp{#1}%
         \else\strut\raise.85ex\hbox{\scriptsize #1}\fi}
\def\sb#1{\ifmmode
           \oldmsb{#1}%
         \else\strut\raise-.54ex\hbox{\scriptsize #1}\fi}
\newbox\@sp
\newbox\@sb
\def\sbp#1#2{\ifmmode%
           \oldmsb{#1}\oldmsp{#2}%
         \else
           \setbox\@sb=\hbox{\sb{#1}}%
           \setbox\@sp=\hbox{\sp{#2}}%
           \rlap{\copy\@sb}\copy\@sp
           \ifdim \wd\@sb >\wd\@sp
             \hskip -\wd\@sp \hskip \wd\@sb
           \fi
        \fi}
\def\msp#1{\ifmmode
           \oldmsp{#1}
         \else \math{\oldmsp{#1}}\fi}
\def\msb#1{\ifmmode
           \oldmsb{#1}
         \else \math{\oldmsb{#1}}\fi}

\def\supon{\catcode`\^=7}

\def\subon{\catcode`\_=8}

\def\supsubon{\supon \subon}

\newcommand\actcharon{\catcode`\~=13}

\newcommand\paramon{\catcode`\#=6}

\comment{And now to turn us totally on and off...}

\newcommand\reservedcharson{ \commenton  \mathshifton  \atabon  \supsubon 
                             \actcharon  \paramon}

\catcode`@=12
\reservedcharson

\if\papercls \apjcls

\else

\fi

\newcommand\tnm[1]{\tablenotemark{#1}}

\newcommand\chisq{\ifmmode{\chi\sp{2}}\else\math{\chi\sp{2}}\fi}
\newcommand\redchisq{\ifmmode{ \chi\sp{2}\sb{\rm red}}
                    \else\math{\chi\sp{2}\sb{\rm red}}\fi}
\newcommand\Teq{\ifmmode{T\sb{\rm eq}}\else$T$\sb{eq}\fi}
\newcommand\mjup{\ifmmode{M\sb{\rm Jup}}\else$M$\sb{Jup}\fi}
\newcommand\rjup{\ifmmode{R\sb{\rm Jup}}\else$R$\sb{Jup}\fi}
\newcommand\msun{\ifmmode{M\sb{\odot}}\else$M\sb{\odot}$\fi}
\newcommand\rsun{\ifmmode{R\sb{\odot}}\else$R\sb{\odot}$\fi}
\newcommand\mearth{\ifmmode{M\sb{\oplus}}\else$M\sb{\oplus}$\fi}
\newcommand\rearth{\ifmmode{R\sb{\oplus}}\else$R\sb{\oplus}$\fi}

\makeatletter
\newcommand{\nHI}{n_{\mbox{\tiny H\Rmnum{1}}}}

\newcommand{\sHI}{\sigma_{\mbox{\tiny H\Rmnum{1}}}}
\newcommand{\betaHI}{\beta_{\mbox{\tiny H\Rmnum{1}}}}
\newcommand{\nuHI}{\nu_{\mbox{\tiny H\Rmnum{1}}}}
\newcommand{\nuHeI}{\nu_{\mbox{\tiny He\Rmnum{1}}}}

\newcommand{\Rmnum}[1]{\expandafter\@slowromancap\romannumeral #1@}
\newcommand{\HI}{{\rm H\,{\scriptstyle I}}}
\newcommand{\HII}{{\rm H\,{\scriptstyle II}}}
\newcommand{\HeII}{{\rm He\,{\scriptstyle II}}}

\newcommand{\OII}{{\rm O\,{\scriptstyle II}}}
\newcommand{\OIII}{{\rm O\,{\scriptstyle III}}}

\newcommand{\OVI}{{\rm O\,{\scriptstyle VI}}}
\newcommand{\NV}{{\rm N\,{\scriptstyle V}}}

\newcommand{\xHI}{x_{\mbox{\tiny H\Rmnum{1}}}}
\newcommand{\xHII}{x_{\mbox{\tiny H\Rmnum{2}}}}
\newcommand{\Lya}{\ifmmode{\mathrm{Ly}\alpha}\else Ly$\alpha$\xspace\fi}
\newcommand{\NHI}{N_{\mbox{\tiny H\Rmnum{1}}}}
\newcommand{\NHILya}{N_{\mbox{\tiny H\Rmnum{1}},\rm Ly\alpha}}
\newcommand{\NH}{N_{\mbox{\tiny H}}}
\newcommand{\sigmaHI}{\sigma_{\mbox{\tiny H\Rmnum{1}}}}

\makeatother

\shorttitle{A physical origin of LyC and Ly$\alpha$ escape}
\shortauthors{Kakiichi \& Gronke}

\begin{document}

\title{Lyman Radiation Hydrodynamics of Turbulent {H\,{\large \bf II}} Regions in Molecular Clouds: \\
 A Physical Origin of LyC Leakage and the Associated Ly$\boldsymbol\alpha$ Spectra}

\author{Koki Kakiichi}
\affiliation{Department of Physics and Astronomy, University College London, UK}

\author{Max Gronke}
\altaffiliation{Hubble fellow}
\affiliation{Department of Physics and Astronomy, University of California, Santa Barbara, USA}

\email{k.kakiichi@ucl.ac.uk}


\begin{abstract}
We examine Lyman continuum (LyC) leakage through $\HII$ regions regulated by turbulence and radiative feedback in a giant molecular cloud in the context of fully-coupled radiation hydrodynamics (RHD). The physical relations of the LyC escape with $\HI$ covering fraction, kinematics, spectral hardness, and the emergent Lyman-$\alpha$ (Ly$\alpha$) line profiles are studied using a series of RHD turbulence simulations performed with \textsc{ramses-rt}. The turbulence-regulated mechanism allows ionizing photons to leak out at early times before the onset of supernova feedback. The LyC photons escape through turbulence-generated low column density channels which are evacuated efficiently by radiative feedback via photoheating-induced shocks across the D-type ionization fronts. Ly$\alpha$ photons funnel through the photoionized channels along the paths of LyC escape, resulting in a diverse Ly$\alpha$ spectral morphology including narrow double-peaked profiles. The Ly$\alpha$ peak separation is controlled by the residual $\HI$ column density of the channels and the line asymmetry correlates with the porosity and multiphase structure of the $\HII$ region. This mechanism through the turbulent $\HII$ regions can naturally reproduce the observed Ly$\alpha$ spectral characteristics of some of LyC-leaking galaxies. This RHD turbulence-origin provides an appealing hypothesis to explain high LyC leakage from very young ($\sim$\,3\,Myr) star-forming galaxies found in the local Universe without need of extreme galactic outflows nor supernova feedback. We discuss the implications of the turbulent $\HII$ regions on other nebular emission lines and a possible observational test with the Magellanic System and local blue compact dwarf galaxies as analogs of reionization-era systems. 
\end{abstract}

\keywords{dark ages, reionization, first stars --- $\HII$ regions --- hydrodynamics --- line: formation --- radiative transfer --- turbulence}

\section{Introduction}

Understanding the physical origin of how ionizing radiation escape through the interstellar medium (ISM) of star-forming galaxies is critical to understanding the sources of reionization. For galaxies to drive $\HI$ reionization, the escape fraction of Lyman continuum (LyC) photons, $f_{\rm esc}^{\rm LyC}$, must be as large as $\sim10-20\%$ at $6\lesssim z\lesssim12$ \citep[e.g.][]{Robertson2015} to be consistent with the UV luminosity function \citep{Bouwens15,Oesch18} and the measure of the Thomson optical depth to the cosmic microwave background \citep{Planck2018}. At the tail end of reionization, galaxies are thought to drive the large-scale UV background fluctuation in the intergalactic medium \citep{Becker18}, and the indirect measure suggests a probable increase of the average escape fraction to $\gtrsim8~\%$ at $z\gtrsim5-6$ \citep{Kakiichi18,Meyer19}. However, little is known about the physical origin of ionizing escape or the cause of the required rise in the ionizing power of galaxies towards the reionization epoch. 

Recent deep {\it Hubble Space Telescope (HST)} imaging campaigns \citep{Siana15,Fletcher18,Oesch18} and ground-based deep spectroscopic searches \citep{Marchi17,Steidel18} have revealed a signature of the LyC leakages along the lines-of-sight of Ly$\alpha$ emitters (LAEs) and Lyman-break galaxies (LBGs), providing a valuable sample with which the origin of LyC escape can be directly studied. These $z\sim2-3$ LyC leaking LAEs are young and associated with intense $[\OIII]$ emission, which suggests spectrally hard stellar population with low metallicity resembling star-forming systems at $6<z<12$ \citep{Nakajima16}. Further detail is provided by the Cosmic Origin Spectrograph on board {\it HST}, which has revealed LyC detection from low redshift dwarf galaxies with high $[\OIII]/[\OII]$ line ratios \citep{Izotov2016,Izotov2018b}. Complementary Ly$\alpha$ and UV-to-optical spectroscopy and spatially-resolved images of low-redshift LAE analogs \citep{Ostlin14,Hayes14,Jaskot14,Henry15} make it possible to examine the inner working of the LyC-leaking systems.

For triggering the leakage of ionizing radiation, a commonly held view is that stellar feedback such as supernova explosions drive galactic outflows, creating low column density channels in the ISM through which LyC photons escape \citep[e.g.][]{Kimm&Cen14,Wise14}. Observationally, however, while outflows are ubiquitous in LyC-leaking galaxies and some with an extreme value \citep{Heckman11,Borthakur14}, their outflow kinematics may not be statistically different from non LyC-leaking systems \citep{Chisholm17,Jaskot17}. Also, the presence of prominent stellar winds P-Cygni $\OVI$ $\lambda1035$ and $\NV$ $\lambda1240$ profiles from massive stars suggests very young ages ($\sim2-3$ Myr) for local LyC-leaking galaxies \citep{Izotov2018b}, indicating that there is little time for supernova explosions to expel gas from the birth cloud. The $\HI$ absorption spectra in the gamma-ray burst afterglow, which traces the direct environment of star forming regions and the ISM at the death of a massive star \citep{Prochaska06,Vreeswijk13}, reveals ubiquitous optically thick gas and a high $\HI$ covering fraction, indicating the low LyC escape fraction of $<1.5~\%$ \citep{Chen07,Fynbo09,Tanvir19}. These seem to challenge the picture that supernova and galactic outflows solely trigger the LyC leakage. Clearly, the physics is far from simple and the observational diversity requires that any successful theory of escape fraction should be able to explain not only why the escape fraction can be high, but also the diversity of LyC escape fractions.

To this end, we wish to examine other mechanisms of LyC leakage invoking turbulence and $\HII$ region feedback that can operate at an early time of the star-forming clouds before supernova explosions occur. The formation and evolution of the $\HII$ region in a turbulent molecular cloud is a natural consequence of the gravoturbulent fragmentation paradigm of star formation \citep[e.g.][]{Krumholz05,Federrath12}, which has been subject to many theoretical studies \citep{Mellema06,Arthur11,Krumholz2006,Krumholz2012,Dale2012,Kim2016,Kim2018}. Recent high-resolution, cosmological galaxy formation simulations suggest that the small 10-100 pc scale environment around star-forming regions is likely the key process in regulating the leakage of ionizing radiation \citep{Kimm&Cen14,Paardekooper15,Trebitsch17}. The galaxies experience substantial LyC leakage when the optically thick gas is evacuated from the parsec-scale environment of massive stars during the period when the stars still can provide abundant ionizing photons \citep{Ma15,Ma16}. The spatial scale required for understanding LyC leakage is indeed approaching that of giant molecular clouds (GMCs).

For any given scenario of LyC leakage, it is critical to understand the connection between the LyC escape fraction and other spectroscopic features including Ly$\alpha$ and $[\OIII]/[\OII]$ line ratio. This is important as a test of a theory. Also, because the direct LyC leakage from $6<z<12$ galaxies at the heart of reionization era cannot be observed even with {\it James Webb Space Telescope (JWST)}, any inference on their ionizing capabilities must rely on an interpretation of other observable rest-UV or optical signatures such as nebular emission \citep{Inoue11,Zackrisson13,Zackrisson17,Tamura18} and UV absorption lines \citep{Jones13,Leethochawalit16,Gazagnes18,Chisholm18}. 

Ly$\alpha$ is particularly important because its unique brightness and omnipresence throughout cosmic time allows us to observe it in a large sample of galaxies \citep[e.g.][]{Ouchi18,Wisotzki18}. In addition, Ly$\alpha$ is a resonant line of neutral hydrogen with a cross section at the line center approximately three orders of magnitudes larger than that of LyC photons connecting Ly$\alpha$ escape processes to LyC ones. Furthermore, since each interaction between a $\HI$ atom and a Ly$\alpha$ photon shifts the photon's frequency, the emergent Ly$\alpha$ spectral shape is indicative of the system's neutral hydrogen distribution and kinematics. All these factors make Ly$\alpha$ observables important diagnostics to be correlated with LyC escape. This natural `Ly$\alpha$-LyC' connection led to theoretical studies \citep[e.g.][]{Verhamme15,Dijkstra16}, and recently the correlation between the Ly$\alpha$ and LyC escape fraction as well as other Ly$\alpha$ line properties, e.g. peak separation, has been observationally studied \citep{Verhamme17,Izotov2018b,Marchi18,Steidel18}.

However, thus far, these Ly$\alpha$-LyC studies have relied on either simplified models of gas and kinematics, e.g. a shell or clumpy medium \citep{Verhamme15,Dijkstra16}, or post-processing of cosmological galaxy formation simulations \citep{Yajima14}. Although there is a substantial progress  both in cosmological \citep[e.g.][]{Gnedin16,Pawlik15,Pawlik17,Rosdahl18} and zoom-in simulations for understanding the origin of Ly$\alpha$ \citep{Smith19} as well as nebular and infared lines \citep{Katz19}, resolving the sub-parsec structures in $\HII$ regions, GMCs, and cold gas phase in general including the circum-galactic medium (CGM) \citep[e.g.][]{vandeVoort19,Hummels2018} still remains difficult. A study of the connection between LyC escape and Ly$\alpha$ transfer using the detailed radiation hydrodynamic (RHD) simulations of  individual $\HII$ regions and GMCs (\citealt{Geen15,Geen16,Howard17,Howard18}; \citealt{Kim2018}) has not yet been carried out (except for \citealt{Kimm19} upon completion of this work). In this paper, we examine the LyC leakage mechanism and the emergent Ly$\alpha$ line profiles using a series of RHD turbulence simulations representing a patch of a $\HII$ region in a GMC. Using the controlled local simulations, we analyze the process responsible for Ly$\alpha$-LyC connection and the relation to turbulence kinematics, radiative feedback, and spectral hardness of ionizing sources. Our goal is to understand the origin of the Ly$\alpha$-LyC connection found in the observed LyC-leaking sample as well as to provide a benchmark for future global simulations of molecular clouds and galaxies.

The paper is organized as follows. Section~\ref{sec:theory} introduces the various regimes of LyC leakage. The numerical simulations and set up are described in Section~\ref{sec:simulation}. We presents the results on the LyC leakage mechanism through the turbulent $\HII$ regions in Section~\ref{sec:result}, followed by the connection with the emergent Ly$\alpha$ line profiles in Section~\ref{sec:Lya-LyC}. We discuss the the limitation of our simulations and possible observational tests in Section~\ref{sec:discussion}. The conclusion is summarized in Section~\ref{sec:conclusion}.

\begin{figure}
\hspace*{-0.3cm}
\includegraphics[width=1.1\columnwidth]{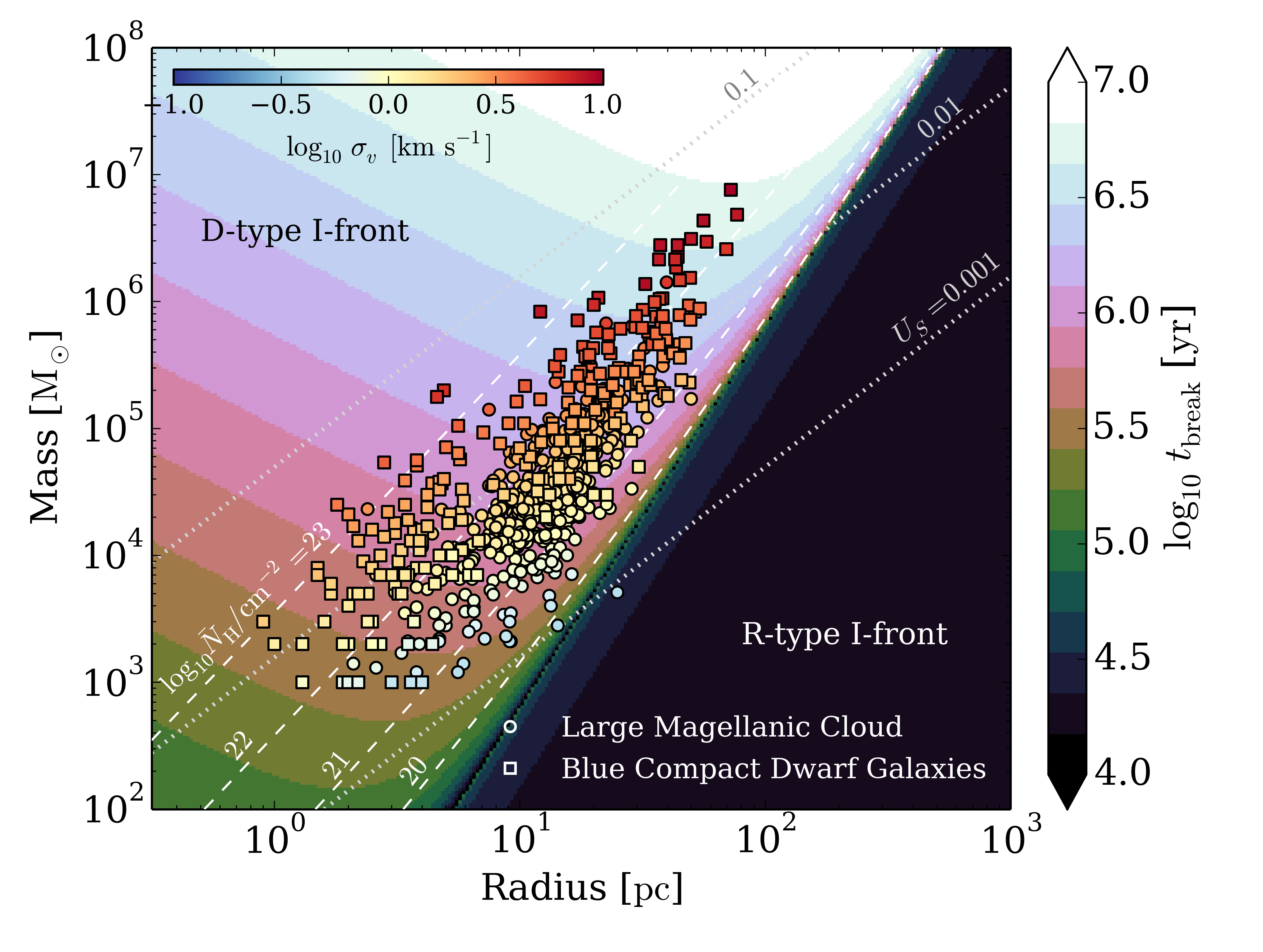}
\vspace{-0.5cm}
\caption{The filled contours show the time required for the I-front to break out of a GMC, $t_{\rm break}$, at a given mass and radius assuming a spherical uniform cloud. The circles and squares denote the observed mass, radius, and velocity dispersion of GMCs in Large Magellanic Cloud \citep{Wong11} and blue compact dwarf galaxies \citep{Kepley16, Miura18, Imara&Faesi18} are overlaid to guide a relevant parameter space. The dashed lines indicate the total gas column density of the shell between the initial Str\"omgren radius $R_S$ and cloud radius $R_{\rm cl}$ and the dotted lines indicates the ionization parameter at the Str\"omgren radius. The figure illustrates that I-fronts in the majority of GMCs are likely the D-type.}\label{fig:GMC}
\end{figure}

\section{LyC Leakage Mechanism}\label{sec:theory}

In the simple commonly-held picture \citep{Zackrisson13,Reddy16,Steidel18}, LyC leakage is leveraged by ({\it i}) ionization-bound with holes or ({\it ii}) density-bound nebulae. The escape fractions in the two regimes are as follows.

\vspace{0.2cm}
{\it Ionization-bound LyC leakage}: In this scenario, the LyC photons escape through holes of low column density channels through the ISM, but in the other directions the ionization front (I-front) is bound within the nebula (i.e ionization-bound). Thus, the escape fraction is determined by the availability of the holes through the ISM,
\begin{equation}
f_{\rm esc}^{\rm LyC}(\mbox{ionization-bound})\sim 1- f_{\rm cov},
\end{equation}
where $f_{\rm cov}$ is the fraction of lines-of-sight around ionizing sources (e.g. massive stars) covered by the optically thick gas. Such holes can be created by turbulence or stellar feedback including photoionization, radiation pressure, stellar winds, and supernova.

\vspace{0.2cm}
{\it Density-bound LyC leakage}: In this scenario, the intense radiation from massive stars ionizes  all the gas in the ISM, thus the I-front is no longer bound inside the system (i.e density-bound). This allows LyC photons to brute-forcefully escape out of the system by ionizing all the gas along the way. Thus, the escape fraction is given by a fraction of photons that have not be absorbed in the ISM \citep{Dove&Shull94,Benson13},  
\begin{equation}
f_{\rm esc}^{\rm LyC}(\mbox{density-bound})\sim 1- \frac{\dot{N}_{\rm rec}}{\dot{N}_{\rm ion}},\label{eq:2}
\end{equation}
where $\dot{N}_{\rm rec}$ is the recombination rate and $\dot{N}_{\rm ion}$ is the LyC photon production rate of the star-forming regions. 

\vspace{0.2cm}
In reality, an individual $\HII$ region will consist of both ionization-bound and density-bound directions and a galaxy consists of an ensemble of the $\HII$ regions. This highlights the two important factors in controlling LyC escape: ({\it i}) the modes of creating a low column density channels, e.g. by turbulence, radiative and/or stellar feedback, through which I-fronts can break out of a natal star-forming cloud and ({\it ii}) the large LyC production rate of star-forming regions to compensate the recombination rate in the density-bound photoionized channels.

\subsection{Nature of I-fronts: \\ $\HII$ Regions in Turbulent Molecular Clouds}

In the gravoturbulent fragmentation paradigm \citep[e.g.][]{Krumholz05,Federrath12} stars form in dense clumps generated by the supersonic turbulence in a GMC, providing a low star formation efficiency $\epsilon_\star$ (a few to several per cent, e.g. \citealt{McKee&Ostriker07,Kennicutt&Evans12}) -- defined as a fraction of cloud's gas mass that is converted into stars -- and kinetic support against gravitational collapse. In Figure \ref{fig:GMC} we show the observed mass, size, and turbulent velocity dispersion of giant molecular clouds in the Large Magellanic Cloud \citep{Wong11} and blue compact dwarf galaxies (II~Zw~40, \citealt{Kepley16}; NGC 5253, \citealt{Miura18}; Henize 2-10, \citealt{Imara&Faesi18}) which share similarity properties to those observed at high redshift \citep{Izotov11,Crowther17}. Although the properties of molecular clouds in LyC-leaking systems and high-redshift galaxies are unknown, GMCs in the environments of nearby dwarfs serve as a guideline for the relevant parameter space. 

To identify the LyC leakage mechanism, let us consider a spherical homogeneous cloud of gas mass $M_{\rm cl}$ and radius $R_{\rm cl}$ with a central source of stellar mass $M_\star=\epsilon_\star M_{\rm cl}$ with an ionizing photon production rate (in units of $\rm photons~s^{-1}$),
\begin{align}
\dot{N}_{\rm ion}&=\xi_{\rm ion}\epsilon_\star M_{\rm cl} \nonumber \\
&\simeq1.9\times10^{50}\left(\frac{\epsilon_\star}{0.05}\right)\left(\frac{M_{\rm cl}}{10^5\rm M_\odot}\right)\rm~s^{-1}.
\end{align}
A single stellar population of $1\rm~Myr$ after starburst with stellar metallicity $Z_\star=0.002$ from binary stellar population code \textsc{bpass} (\textsc{bpassv2.1\_imf135\_100}, \citealt{Eldridge17}) yields the ionizing photon production rate per stellar mass of $\xi_{\rm ion}=3.8\times10^{46}\rm~ph~s^{-1}~M_\odot^{-1}$. The corresponding Str\"omgren radius is 
\begin{align}
R_S&=\left(\frac{3\dot{N}_{\rm ion}}{4\pi a_{\rm B}\bar{n}_0^2}\right)^{1/3} \nonumber \\ 
&\simeq7.4\left(\frac{\epsilon_\star}{0.05}\frac{M_{\rm cl}}{10^5\rm M_\odot}\right)^{1/3}\mspace{-5mu}\left(\frac{R_{\rm cl}}{20\rm pc}\right)^{2}\rm~pc,
\end{align}
where $\alpha_{\rm B}\approx2.6\times10^{-13}T_4^{-0.7}\rm cm^3~s^{-1}$ ($T_4=T/10^4\rm~K$) is the case B recombination rate  and $\bar{n}_0\simeq120.8\times\left(M_{\rm cl}/10^5\rm~M_\odot \right)\left( R_{\rm cl}/20\rm~pc \right)^{-3}\rm cm^{-3}$ is the mean number density of hydrogen nuclei. The dimensionless ratio of the cloud radius to the Str\"omgren radius sets whether the $\HII$ region can break out of the cloud during its early phase of rapid expansion (known as R-type). If $R_{\rm cl}<R_S$, the R-type I-front radius expands as $r_{\rm I}(t)=R_S(1-e^{-t/t_{\rm rec}})^{1/3}$ \citep[e.g.][]{Shu1992} and approaches rapidly to the Str\"omgren radius in the order of recombination timescale $t_{\rm rec}=(\alpha_B\bar{n}_0)^{-1}\approx1.2 T_4^{0.7}(\bar{n}_0/100\rm~cm^{-3})^{-1}\rm~kyr$. The R-type I-front therefore breaks out of the parent cloud, $r_{\rm I}(t_{\rm break})=R_{\rm cl}$, after
\begin{equation}
t_{\rm break}=t_{\rm rec}\ln\left[1-\left(\frac{R_{\rm cl}}{R_S}\right)^3\right]^{-1} \mbox{for R-type I-front}.
\end{equation}
The LyC leakage immediately follows the density-bound regime. Using Equation (\ref{eq:2}), we find
\begin{equation}
f_{\rm esc}^{\rm LyC}=1-\frac{\alpha_{\rm B}\bar{n}_0}{m_{\rm H}\epsilon_\star\xi_{\rm ion}}~\mbox{after R-type breakout}.
\end{equation}
The density-bound LyC leakage following the early R-type I-front only occur for a diffuse GMC (in a spherical homogeneous cloud model); for example, a cloud with $M_{\rm cl}=10^5\rm~M_\odot$ and $R_{\rm cl}=60\rm~pc$ gives $\bar{n}_0\approx4.5\rm~cm^{-3}$ and $f_{\rm esc}^{\rm LyC}\approx0.72$ in the case of the R-type I-front.

In a realistic dynamical $\HII$ region where there is a large temperature contrast between the photoheated $\HII$ region and the ambient cold neutral gas, this produces a shock front ahead of the I-front which pushes the gas outwards, enabling the I-front (known as D-type) to proceed beyond the Str\"omgren radius \citep[e.g.][]{Whalen04,Krumholz07}. This allows a dynamical transition from the initially ionization-bound nebula to the density-bound regime at later time. The D-type I-front expands as $r_{\rm I}(t)\approx R_S\left[1+7 c_{{\rm\scriptscriptstyle II}}t/(4R_S)\right]^{4/7}$ \citep[see][Chapter 20]{Shu1992} with the velocity of the order of sound speed of ionized gas $c_{{\rm II}}=\sqrt{2k_{\rm B}T/m_{\rm H}}=12.8T_4^{1/2}\rm~km~s^{-1}$ and breaks out of the cloud after 
\begin{equation}
t_{\rm break}=\frac{4R_S}{7c_{{\rm II}}}\left[\left(\frac{R_{\rm cl}}{R_S}\right)^{7/4}-1\right]~\mbox{for D-type I-front}.
\end{equation}
After the breakout of the D-type I-front, because the gas is evacuated by the thermal pressure, the interior density is lowered to $\bar{n}_{\rm II}=(R_S/r_{\rm I})^{3/2} \bar{n}_0$ \citep{Shu1992}. Thus, the resulting density-bound LyC leakage is increased as there is less gas in the $\HII$ region,
\begin{equation}
f_{\rm esc}^{\rm LyC}=1-\frac{\alpha_{\rm B}\bar{n}_0}{m_{\rm H}\epsilon_\star\xi_{\rm ion}}\left(\frac{\bar{n}_{\rm II}}{\bar{n}_0}\right)^2~\mbox{after D-type breakout}.
\end{equation}
This gives a high LyC escape fraction for an initially ionization-bound nebula by the action of photoionization heating and the associated I-front shocks.

In Figure \ref{fig:GMC} we overlay the estimated I-front breakout time for each parameter space of molecular clouds. For most of the observed molecular clouds, $\HII$ regions follow the D-type I-front. More luminous and massive GMCs that contribute to the large fraction of the total ionizing photon budget of a galaxy require longer times for the D-type I-front to break out, which must compete with the short $\sim$Myr lifetime of massive stars. This means that the LyC leakage from the dynamical $\HII$ region in a molecular cloud must be treated with fully-coupled radiation hydrodynamics. Typical ionization parameters at the Str\"omgren radius and total hydrogen column density between the Str\"omgren radius and cloud radius are approximately  $U_S=\dot{N}_{\rm ion}/(4\pi R_S^2 \bar{n}_0c)\sim0.01$ and $\bar{N}_{\rm H}\sim10^{22}\rm~cm^{-2}$. The turbulent velocities of the massive GMCs of $M_{\rm cl}\sim10^{5-6}\rm~M_{\odot}$ are $\sigma_v\sim 1-10\rm~km~s^{-1}$. The turbulent nature of the GMCs clearly introduces anisotropy and inhomogeneity to this simple back-of-envelope view. Thus, having identified the relevant regime of LyC leakage and the approximate parameter space of interest, we present a detailed account of LyC leakage through a patch of a turbulent molecular cloud in Section~\ref{sec:simulation}. 

\begin{figure}
\vspace{0.2cm}
\includegraphics[width=\columnwidth]{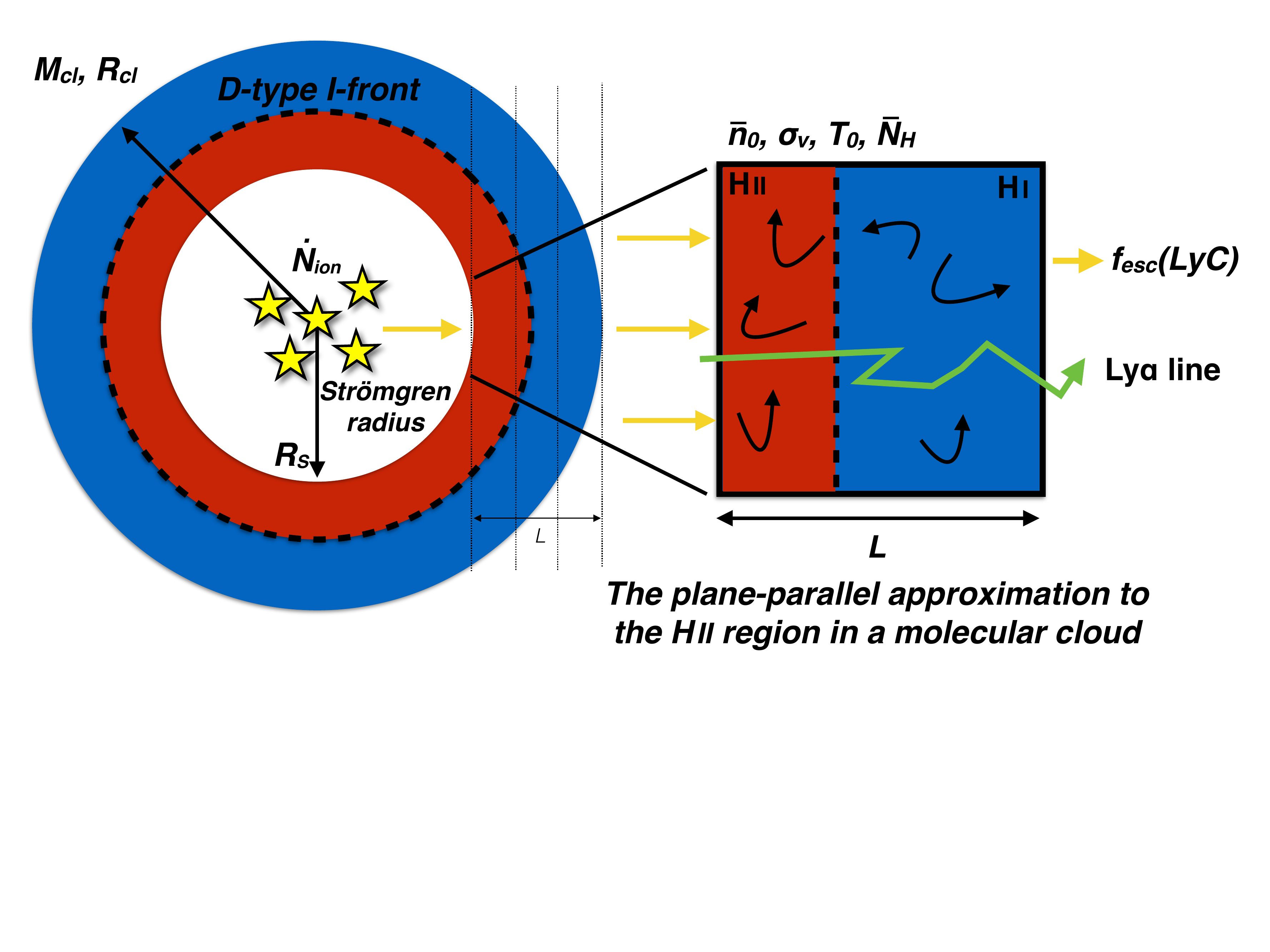}
\vspace{-2.2cm}
\caption{Schematic illustration of the simulation setup. Our simulation boxes represent a patch of the $\HII$ region in a GMC irradiated by a single stellar population with an ionizing luminosity $\dot{N}_{\rm ion}$. Initially the R-type I-front propagates rapidly to the Str\"omgren radius $R_S$ and then the $\HII$ region gradually grows by the D-type I-front. A patch of size $L$ located at a distance $d$ from the stellar source is characterized by the initial mean density $\bar{n}_0$, temperature $T_0$, rms turbulent velocity $\sigma_v$, and the total gas column density $\bar{N}_{\rm H}$. }\label{fig:cartoon}
\end{figure}

\section{Physical Formulation and Simulations}\label{sec:simulation}

\subsection{Equations of Radiation Hydrodynamics}

We now turn from a heuristic argument to a physical formulation of LyC leakage and the associated Ly$\alpha$ RT in a full radiation hydrodynamical framework. We consider a plane-parallel atmosphere (slab) of turbulent gas cloud around a star-forming region with an initial number density $\bar{n}_0$ and total hydrogen column density $\bar{N}_{{\rm \tiny H},0}$ (and size $L=\bar{N}_{{\rm \tiny H},0}/\bar{n}_0$) which is continuously irradiated by the ionizing radiation from a star-forming cluster (see Figure \ref{fig:cartoon}). For simplicity we have assumed a hydrogen-only gas. The system is constantly perturbed on the large scale to maintain the turbulence to represent a large-scale forcing such as gas accretion or disk instability in a galaxy \citep{Elmegreen10,Goldbaum11,Krumholz16}. This setup was previously employed by \citet{Gritschneder09,Gritschneder10}. 

The fundamental equations of radiation hydrodynamics that govern the distribution and kinematics of the gas and the transport of ionizing radiation are,
\begin{align}
&\mspace{-10mu}
\frac{\partial\rho}{\partial t}+\nabla\cdot(\rho \boldsymbol v )=0 
\\
&\mspace{-10mu}
\frac{\partial\rho\boldsymbol{v}}{\partial t}
+\boldsymbol\nabla\cdot(\rho\boldsymbol{vv})=-\boldsymbol\nabla P+\rho\boldsymbol{f}_{\rm stir}+\rho \boldsymbol{f}_{\rm rad}
\label{eq:momentum} \\
&\mspace{-10mu}
\frac{\partial E}{\partial t}+\boldsymbol\nabla\cdot[(E+P)\boldsymbol{v}]= \rho\boldsymbol{v}\cdot\boldsymbol{f}_{\rm stir}+\rho\boldsymbol{v}\cdot\boldsymbol{f}_{\rm rad}+\Lambda 
\label{eq:energy} \\
&\mspace{-10mu}
\frac{1}{c}\frac{\partial E_\nu}{\partial t}+\boldsymbol\nabla\cdot\boldsymbol{F}_\nu=-\nHI \sHI E_\nu+F^{\rm inc}_\nu\delta_D({\boldsymbol r}-{\boldsymbol r}_0)
\\
&\mspace{-10mu}
\frac{1}{c}\frac{\partial \boldsymbol{F}_\nu}{\partial t}+c\boldsymbol\nabla\cdot(\boldsymbol{\mathsf{f}}_{\nu} E_{\nu})=-\nHI \sHI\boldsymbol{F}_\nu
\end{align}
which couples to the rate equation,
\begin{equation}
\frac{d\nHI}{dt}=\alpha_B n_p n_e-(\Gamma+\betaHI n_e)\nHI,~\Gamma=\int \sHI\frac{F_\nu}{h\nu}  d\nu
\end{equation}
where $\rho$, $\boldsymbol v$, $P$, and $E$ are the density, velocity, thermal pressure, and total (thermal plus kinetic) energy density of the gas, $E_\nu$ and $\boldsymbol{F}_\nu$ ($F_\nu=|\boldsymbol{F}_\nu|$) are the specific energy density and and flux of the ionizing radiation, $\boldsymbol{\mathsf{f}}_\nu$ is the Eddington tensor, $\nHI$ is the number density of neutral hydrogen, $\sHI=\sigma_L(\nu/\nu_L)^{-3}$ is the photoionization cross-section of atomic hydrogen ($\sigma_L=6.3\times10^{-18}\rm~cm^2$ and $h\nu_L=13.6\rm~eV$), $\betaHI(T)$ is the collisional ionization rate coefficient, and the force exerted by the ionizing radiation pressure is
\begin{equation}
{\boldsymbol f}_{\rm rad}=\frac{\nHI}{\rho c}\int\sHI\boldsymbol{F}_\nu d\nu. 
\end{equation}

The heating and cooling are treated approximately. $\Lambda=\mathcal{H}+\mathcal{L}$ is the sum of the rates of radiative heating $\mathcal{H}$ and cooling $\mathcal{L}$ in units of energy per unit time per unit volume. The heating term includes the $\HI$ photoionization heating and the cooling term includes recombination, collisional ionization and excitation, Bremsstrahlung coolings \citep{Rosdahl13}. We do not explicitly follow any metal line cooling nor other heating/cooling mechanisms. Instead, following \citet{Gritschneder09}, we assume an isothermal equation of state (adiabatic index of $\gamma=1$) to approximate complex thermal exchange mechanism such that adiabatic compression and expansion retain the isothermality of the gas.\footnote{In this way, the neutral gas ahead of I-front remains at the isothermal initial temperature $T_0~(=100\rm~K)$, and the photoionized gas retains an approximate isothermality at $\sim10^4\rm~K$.} 

The external random force field $\boldsymbol{f}_{\rm stir}$ is applied to excite turbulent flow \citep[e.g.][]{Kritsuk07,Federrath10}.  We assume a Gaussian random field with a flat power spectrum with power only in the first four largest Fourier modes. We apply a Helmholtz decomposition to the field to produce a different mode (solenoidal or compressive) of large-scale forcing depending on its physical origin, which is parametrized by the forcing parameter $\zeta$ ($\zeta=1$ for purely solenoidal and $\zeta=0$ for purely compressive). The amplitude of the forcing field is chosen to maintain an rms velocity dispersion of a turbulence of interest. The detail is described in Appendix A. 

The incident spectrum $F_\nu^{\rm inc}$ from a star-forming region is produced by a starburst based on the binary stellar population synthesis code \textsc{bpass} \citep{Eldridge17}. The important dimensionless number is the ionization parameter at the incident face
\begin{equation}
\mathcal{U}=\frac{F_{\rm ion}^{\rm inc}}{\bar{n}_0 c}=\frac{1}{\bar{n}_0 c}\int_{\nu_L} \frac{F_\nu^{\rm inc} d\nu}{h\nu}
\end{equation} 
For example, at the ionization parameter $\mathcal{U}=1.3\times10^{-2}$ and $\bar{n}_0=500\rm~cm^{-3}$, the incident flux is $F_{\rm ion}^{\rm inc}=2\times10^{11}\rm~s^{-1}~cm^{-2}$. This corresponds to a slab located at $\simeq 2-10\rm~pc$ away from a single stellar population with the ionizing photon production rate $\simeq10^{50}-10^{51}\rm~s^{-1}$. These values roughly match with those found in global simulations of $\sim 10^4-10^6\rm~M_\odot$ GMCs \citep[e.g.][]{Geen15,Geen16,Kim2018}.

Using the line-of-sight $\HI$ column densities $\NHI$ measured from the outcoming face of the slab, the LyC leakage is defined as the transmitted fraction of ionizing photons with frequency $\nu$,
\begin{equation}
\mathcal{T}_{\rm LyC}(\nu)=\int_0^\infty e^{-\sHI\NHI} d\NHI,\label{eq:leakage}
\end{equation} 
and by integrating over all photons escaping from the slab, the LyC escape fraction is given by
\begin{equation}
f_{\rm esc}^{\rm LyC}=\frac{\displaystyle\int_{\nuHI}^{\nuHeI} \mathcal{T}_{\rm LyC}(\nu)\frac{F_{\nu}^{\rm inc}}{h\nu}d\nu}{\displaystyle\int_{\nuHI}^{\nuHeI} \frac{F_{\nu}^{\rm inc}}{h\nu}d\nu}.\label{eq:escape}
\end{equation}
Note that the escape fraction can also be measured by directly taking the ratio between incoming and outcoming fluxes. Both $\NHI$-based and flux-based estimators agree well \citep{Trebitsch17}. We also verified that our frequency-dependent definition gives consistent results with the frequency-integrated RHD simulations. We use Equations (\ref{eq:leakage}) and (\ref{eq:escape}) to measure the LyC escape fraction throughout this paper.

\begin{deluxetable*}{lllllllll}
\tabletypesize{\footnotesize}
\tablecaption{Simulation setup\,\tnm{(a)} \label{table:setup}}
\tablehead{\colhead{Name} & $\bar{n}_0$ & $\sigma_v$ & $F_{\rm ion}^{\rm inc}$ & $\bar{N}_{{\rm \tiny H},0}$ &  $\mathcal{U}$ & $\mathcal{M}_{{\rm I},0}$ & $\zeta$ & Comment \\
   & [$\rm cm^{-3}$] & [$\rm km~s^{-1}$] & [$\rm ph~s^{-1}~cm^{-2}$] & [$\rm cm^{-2}$] & & & & }
\startdata
\texttt{V18S\_f2e11\_RHD}  & $500$ & $18$ & $2\times10^{11}$ & $7.7\times10^{21}$ & $0.013$ & $20$ & $1$ & fiducial RHD run \\
\texttt{V9S\_f2e11\_RHD}   & $500$ & $9$  & $2\times10^{11}$ & $7.7\times10^{21}$ & $0.013$ & $10$ & $1$ & turbulence series \\
\texttt{V2S\_f2e11\_RHD}   & $500$ & $2$  & $2\times10^{11}$ & $7.7\times10^{21}$   & $0.013$ & $2$ & $1$ & turbulence series \\
\texttt{V18S\_f2e10\_RHD}  & $500$ & $18$ & $2\times10^{10}$ & $7.7\times10^{21}$   & $0.0013$& $20$ & $1$ & spectral hardness series \\
\texttt{V18S\_f8e10\_RHD}  & $500$ & $18$ & $8\times10^{10}$ & $7.7\times10^{21}$   & $0.0052$& $20$ & $1$ & spectral hardness series\,\tnm{(b)} \\
\texttt{V18S\_f1e11\_RHD}  & $500$ & $18$ & $1\times10^{11}$ & $7.7\times10^{21}$   & $0.0065$ & $20$ & $1$ & spectral hardness series \\
\texttt{V18S\_f4e11\_RHD}  & $500$ & $18$ & $4\times10^{11}$ & $7.7\times10^{21}$   & $0.026$ & $20$ & $1$ & spectral hardness series \\
\texttt{V18S\_f8e11\_RHD}  & $500$ & $18$ & $8\times10^{11}$ & $7.7\times10^{21}$   & $0.052$ & $20$ & $1$ & spectral hardness series \\
\texttt{V18C\_f2e11\_RHD}  & $500$ & $18$ & $2\times10^{11}$ & $7.7\times10^{21}$   & $0.013$ & $20$ & $0$ & compressive forcing \\
\texttt{V18S\_f2e11\_RT}   & $500$ & $18$ & $2\times10^{11}$ & $7.7\times10^{21}$   & $0.013$ & $20$ & $1$ & post-processed RT\,\tnm{(c)} \\
\texttt{V18S\_f2e11\_RHD-RP} & $500$ & $18$ & $2\times10^{11}$ & $7.7\times10^{21}$ & $0.013$ & $20$ & $1$ & RHD without radiation pressure\,\tnm{(d)} \\
\enddata
\tablenotetext{(a)}{~~~~For all runs the size of the simulation box is $L=5\rm~pc$ and the initial gas temperature is $100\rm~K$.}
\vspace{-0.1cm}
\tablenotetext{(b)}{~~~~For a comparison we also run a low velocity dispersion simulation with $\sigma_v=2\rm~km~s^{-1}$, keeping all the other parameters the same. }
\vspace{-0.1cm}
\tablenotetext{(c)}{~~~~The post-processed RT solves only the radiative transfer equations keeping the density and velocity fields static fixed at the initial condition. This effectively corresponds to setting $\boldsymbol{\nabla}P=0$, $\boldsymbol{f}_{\rm stir}=0$, and   $\boldsymbol{f}_{\rm rad}=0$ in Equations (\ref{eq:momentum}) and (\ref{eq:energy}). }
\vspace{-0.1cm}
\tablenotetext{(d)}{~~~~The radiation pressure is switched off by setting $\boldsymbol{f}_{\rm rad}=0$ in Equations (\ref{eq:momentum}) and (\ref{eq:energy}). }
\vspace{-0.5cm}
\end{deluxetable*}

\subsection{RHD Turbulence Simulations}

We simulate the above problem using \textsc{ramses-rt} \citep{Teyssier02,Rosdahl13}. \textsc{ramses-rt} employs a second-order Godnouv method to solve an Eulerian fully-coupled radiation hydrodynamics on an adaptive mesh refinement grid, and the radiative transfer is solved by the moment method. We use a static uniform grid with $128^3$ resolution with $L=5\rm~pc$ box on a side. We use the MUSCL scheme with HLLC solver for hydrodynamics with the MinMod slope limiter and the Courant timestep factor 0.1. For the isothermal equation of state, we set the adiabatic index close to unity $\gamma\approx1$ to avoid division by zero. For radiative transport, the HLL solver is used to accurately track the shadowing behind dense gas. We use a single frequency group integrated over $13.6<h\nu<24.6\rm~eV$, the M1 closure for the Eddington tensor, and the on-the-spot approximation. We used the reduced speed of light approximation with $10^{-3}c$ to avoid prohibitively long time integration.  

To perform the RHD turbulence simulations, we have first generated initial conditions by running isothermal turbulence simulations without radiative transfer. We set initial density $\bar{n}_0=500\rm~cm^{-3}$ and temperature $T_0=100\rm~K$ with the periodic boundary condition at all faces. In order to drive turbulence, we perturb the flow with a Gaussian random field with power only at the large scales following the method of \citet{Robertson&Goldreich12,Robertson&Goldreich18} (see Appendix A) with an appropriate choice of the forcing parameter; we set $\zeta=1$ for a fiducial run. We evolved the system for a few tens of Eddy turnover times $\sim10T_{\rm eddy}$ where $T_{\rm eddy}=L/(2\sigma_v)$ to ensure the statistical steady state is reached. We then use the snapshot after $2T_{\rm eddy}$ time as an initial condition for the corresponding RHD turbulence simulation. 

Using the initial condition, we then restart the simulation with full RHD. Both turbulence driving and radiative transfer are activated. The box is irradiated from the left boundary with an ionizing flux $F_{\rm ion}^{\rm inc}=1-4\times10^{11}\rm~s^{-1}~cm^{-2}$ (the corresponding ionization parameters are shown in Table \ref{table:setup}). We use the spectrum-integrated cross section corresponding to the spectral shape of the incoming ionizing radiation consistent with the \textsc{bpass} stellar population synthesis code. We set a reflective boundary condition at the left boundary face and outflow boundary condition at the right boundary face, but otherwise periodic boundary condition. We then evolve the system for $2~\rm Myr$. In order to investigate the conditions for LyC leakage, we have varied the simulation setup and parameters which are summarised in Table \ref{table:setup}. 

\vspace{0.5cm}
\subsection{Monte-Carlo Ly$\alpha$ Radiative Transfer}

We  employ the Monte Carlo radiative transfer (RT) code \textsc{tlac} \citep{Gronke14}. Monte Carlo radiative transfer codes track individual photon packages on their trajectory while simultaneously keeping track of their frequency. This includes the change of direction, and the shift in frequency (mostly) due to Doppler boosting during a scattering event (see, e.g. \citealt{Dijkstra17}).

As input we used the simulated $\HI$ number density, temperature, and velocities on the Cartesian grid with the spatial resolution equal to the \textsc{ramses-rt} runs. We inject Ly$\alpha$ photons at the line centre from the left  boundary of the box ($x=0$). We employed $\sim10^4$ photon packets with a dynamical core-skipping scheme \citep{Smith15}. The emergent Ly$\alpha$ line profile is composed of the frequencies of the photons escaping in the positive $x$ direction. To include the back-scatterings we have mirrored the structure around the $x=0$ axis. Thus, in practice the Ly$\alpha$ RT is done on the $256\times128\times128$ grid with a source at the $x=0$ plane. For $y$ and $z$ boundaries, we used a periodic boundary condition. Therefore, our simulated geometry corresponds to that of a semi-inifinite slab \citep{Neufeld90}.

\begin{figure*}
\vspace{0.5cm}
\hspace*{-0.22cm}
\includegraphics[width=1.03\textwidth]{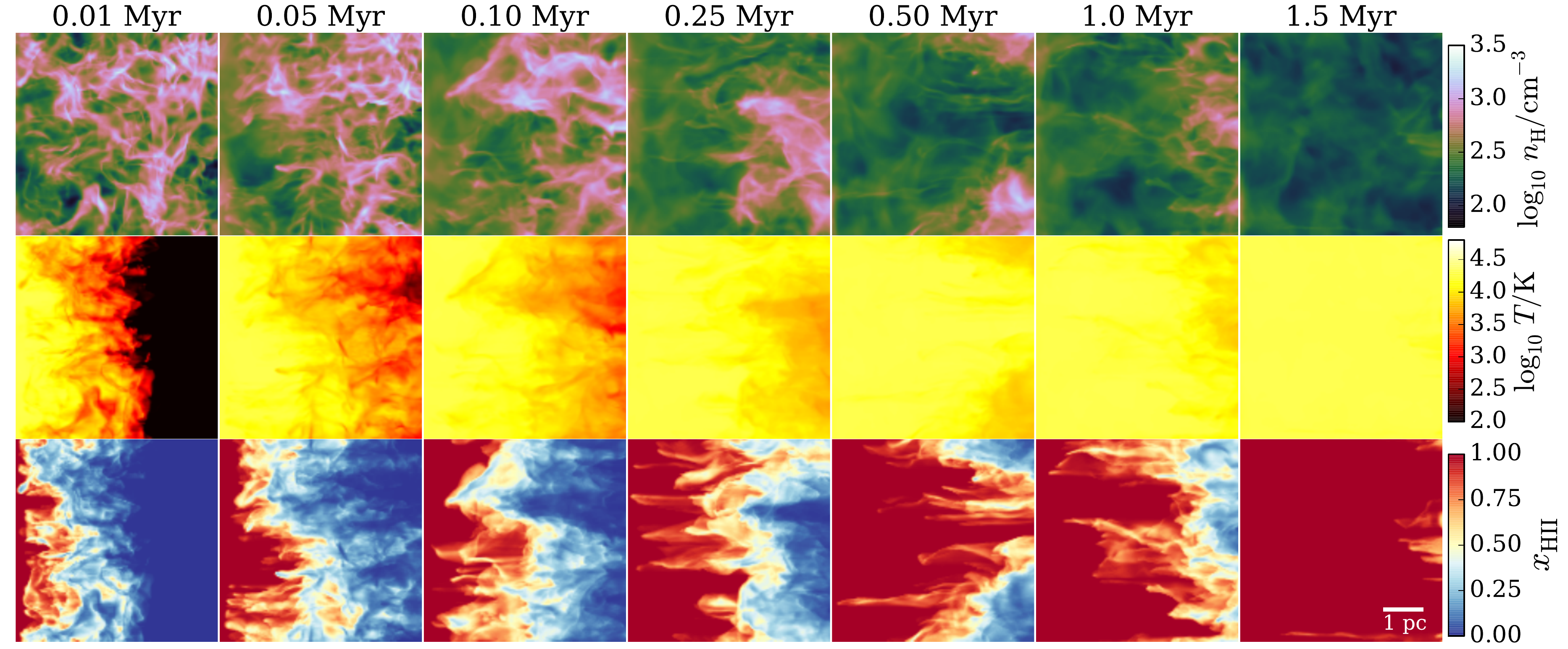}
\caption{Projected maps (side view) of gas density $n_{\rm H}$, mass-weighted temperature $T$ and ionized fraction $\xHII$ as a function of time from left to right (shown is the run  \texttt{V18S\_f2e11\_RHD}). The map is 5 pc per side and the 1 pc length is indicated at the bottom right corner. The ionizing radiation is coming from the left boundary in all panels. An animated version of this figure is downloadable from \href{http://www.star.ucl.ac.uk/~kakiichi/}{http://www.star.ucl.ac.uk/\textasciitilde kakiichi/}.}\label{fig:map}
\end{figure*}

\begin{figure*}
\vspace{-0.7cm}
\hspace*{-0.4cm}
\includegraphics[width=1.03\textwidth]{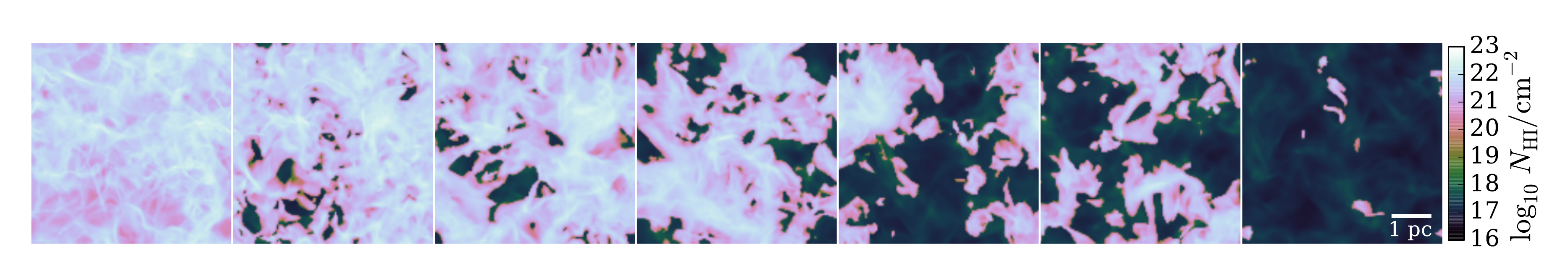}
\vspace{-0.7cm}
\caption{Projected maps (top view) of $\HI$ column density $\NHI$ as a function of time from left to right (as in Figure \ref{fig:map} the snapshots at $t=0.1,\,0.05,\,0.10,\,0.25,\,0.50,\,1.0,\,1.5$ Myr of \texttt{V18S\_f2e11\_RHD} are shown). The viewing direction is towards the non-periodic outcoming face.}\label{fig:NHImap}
\end{figure*}

\begin{figure}
\hspace*{-0.4cm}
\includegraphics[width=1.08\columnwidth]{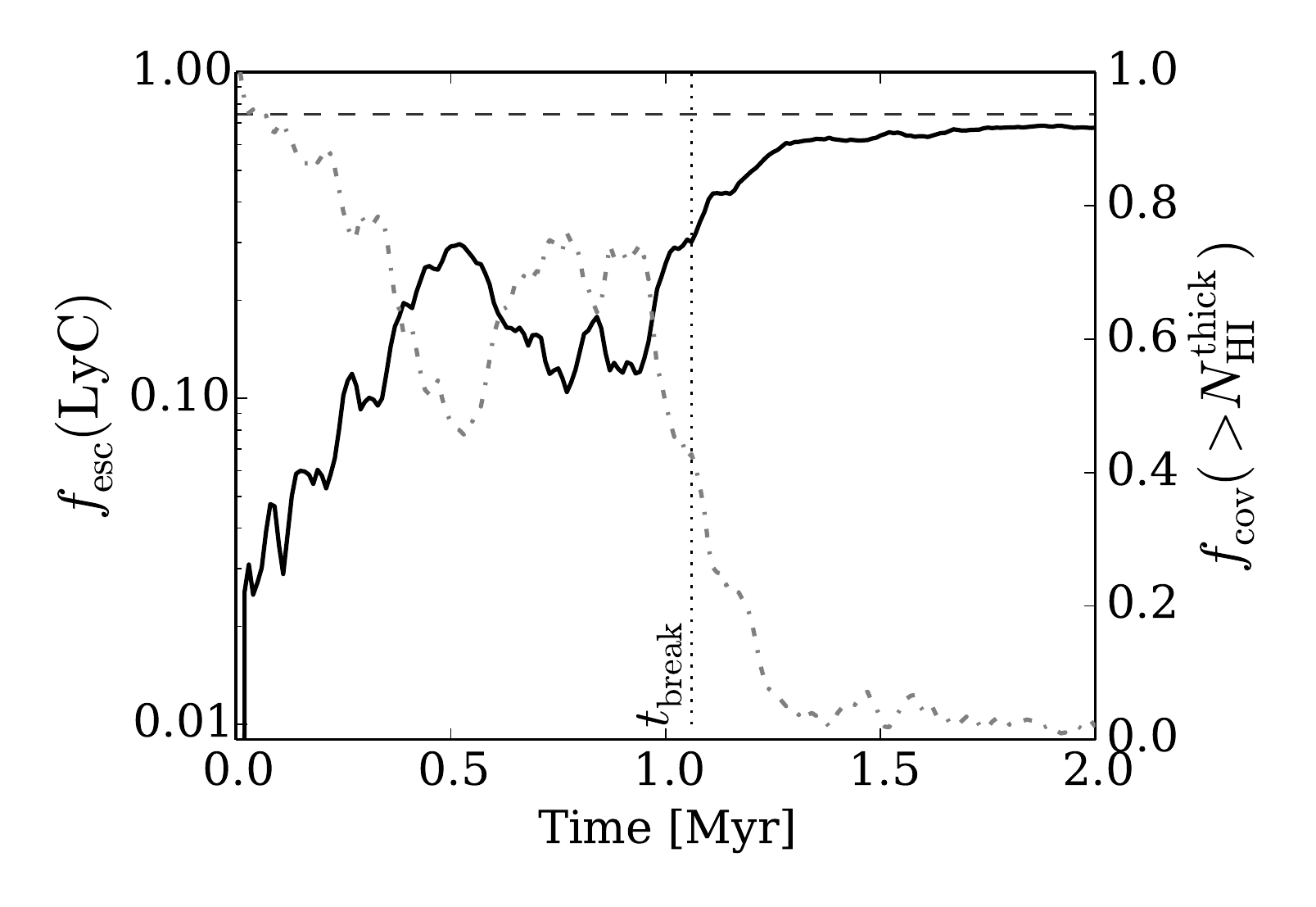}
\vspace{-0.8cm}
\caption{The time evolution of the LyC escape fraction (solid, left y-axis) and covering fraction of optically thick gas $\NHI>7.3\times10^{17}\rm~cm^{-2}$ (dash-dotted, right y-axis) in \texttt{V18S\_f2e11\_RHD} run. The vertical dotted line indicates the I-front breakout time $t_{\rm break}\approx1.1\rm~Myr$. The asymptotic analytic limit of the LyC escape fraction for a fully density-bound nebula, Equation (\ref{eq:22}), is indicated by the horizontal dashed line.}\label{fig:fesc_time}
\end{figure}

\section{Results}\label{sec:result}

\subsection{LyC Escape Fraction}

Here we present an overview of the LyC leakage from a (patch of) $\HII$ region in turbulent molecular clouds. In Figure \ref{fig:map} we show the time evolution of the RHD turbulence. 

Ionizing radiation propagates anisotropically. The photons race ahead in the directions of low column densities opened up by the turbulent fluctuations. The D-type I-fronts do so by creating shocks by the photoheating across the $\HII$ region and the ambient neutral medium, which evacuate the gas effectively in the low column density channels. Figure~\ref{fig:NHImap} confirms the presence of channels of leaking radiation. As supersonic turbulence changes its structure faster than the speed of the D-type I-front of the order of the sound speed of the ionized gas, the low column density channels are not opened sufficiently long enough to sustain the escape of ionizing radiation in the same directions. Instead, the system changes the structure of escaping channels over the Eddy turnover timescale $T_{\rm eddy}\approx245(L/{5\rm~pc})(\sigma_v/10\rm~km~s^{-1})^{-1}\rm~kyr$. The  I-fronts drastically slow down or even halt as soon as dense clumps and filaments are created ahead of them by supersonic shocks in turbulence. That is, {\it LyC photons in a driven turbulent medium need to propagate through a dynamic, constantly changing `maze' before leaking out of the system.} 

These turbulent fluctuations introduce the stochastic variability in the escape fraction on the $\sim100\rm~kyr$ timescale. The time evolution of the escape fraction is shown in Figure \ref{fig:fesc_time}. For example the peak at $\simeq0.5\rm~Myr$ corresponds to the timing of the opening of a large channel (cf. Figure~\ref{fig:map}). Note that this turbulent variability is smaller than the longer $\sim10\rm~Myr$ timescale variability associated with the supernova feedback that exhibits a large $f_{\rm esc}^{\rm LyC}$ variation as the inactive phase of the feedback can completely shut off the leakage. 

While the turbulent fluctuations allow LyC photons to leak out at an early time, the timing at which the breakout of the average I-fronts occurs is delayed. In simulations we define the breakout time as a time when more than 95\% of the entire medium is ionized. This gives $t_{\rm break}\approx1.1\rm~Myr$ for the fiducial RHD run. For a homogeneous slab, the breakout time of the D-type I-front can be computed analytically,
\begin{equation}
t_{\rm break}=\frac{4}{5}\frac{c}{c_{\rm II}}t_{\rm rec}\mathcal{U}\left[\left(\frac{\alpha_{\rm B}N_{\rm H}}{c~\mathcal{U}}\right)^{5/4}-1\right]\sim0.6\rm~Myr,
\end{equation}
for the same parameters used in the fiducial run. Evidently, the average I-front breakout time is delayed for a turbulent medium. There are two reasons for this delay in the breakout time. When the rms turbulent velocity in the $\HII$ region remains supersonic ($\sigma_v>c_{\rm II}\approx12.8T_4^{1/2}\rm~km~s^{-1}$), the density fluctuations can enhance the recombination rate, causing the slow down of the average I-fronts. In addition, the I-front shock-turbulence interaction transports the warm neutral gas ahead of the I-front. This increases the thermal pressure of the ambient gas into which D-type I-front is propagating. This thermal and additional turbulent ram pressure may also contribute to the slow down of the average speed of the D-type I-front \citep{Tremblin14,Geen15}. 

After the breakout $t>t_{\rm break}$, LyC leakage is regulated by the balance between the recombination rate in the $\HII$ region and the incident ionizing flux from the source, approaching to the asymptotic value set by the density-bound regime. The time variability settles down as the medium becomes fully ionized. Since the mechanism of escape is different before and after the breakout that is either dominated by the ionization-bound or density-bound LyC leakage, it is convenient to understand the leaking mechanism in the units of the breakout time. We follow this convention in the rest of the paper.

\begin{figure}
\vspace{0.4cm}
\hspace{-0.4cm}
\includegraphics[width=1.05\columnwidth]{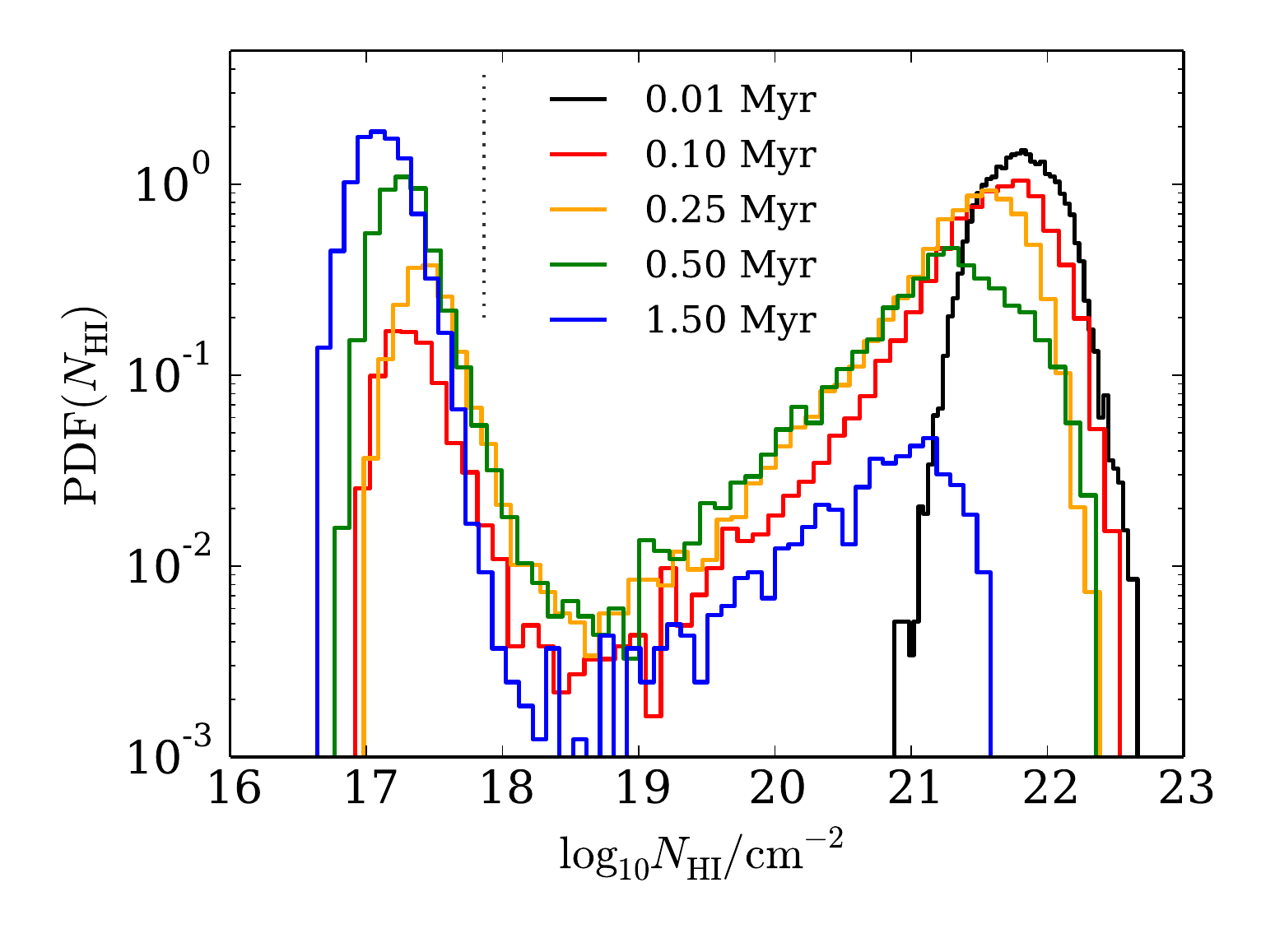}
\vspace{-0.8cm}
\caption{The probability distribution function of the $\HI$ column densities, $\NHI$, at $0.01, 0.10,0.25,0.50,1.50~\rm Myr$ in the \texttt{V18S\_f2e11\_RHD} run. The column density at which a channel becomes optically thick with less than 1\% LyC leakage at Lyman limit is indicated by the vertical dotted line.}\label{fig:NHI}
\end{figure}

\begin{figure}
\hspace*{-0.5cm}
\includegraphics[width=1.1\columnwidth]{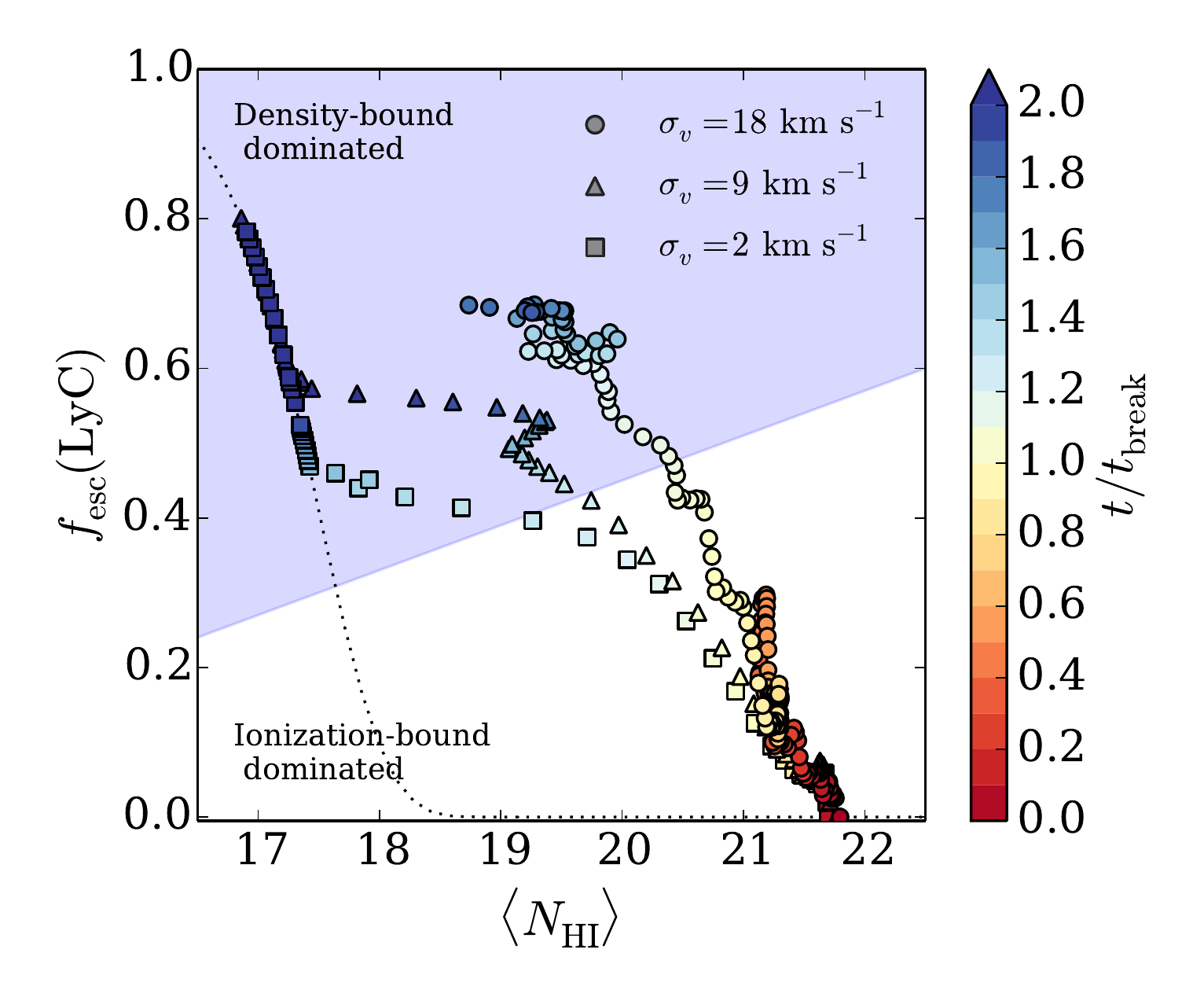}
\vspace{-0.8cm}
\caption{Correlation between LyC escape fraction and average $\HI$ column density $\langle\NHI\rangle$ of the $\HII$ regions. The results from three runs (squares: \texttt{V2S\_f2e11\_RHD}, triangles: \texttt{V9S\_f2e11\_RHD}, circles: \texttt{V18S\_f2e11\_RHD}) are shown. The colors indicate the time normalized by the I-front breakout time of each simulation, $t/t_{\rm break}$. The dotted line indicates the escape fractions for homogeneous media at each $\langle\NHI\rangle$, $f_{\rm esc}^{\rm LyC}=\frac{\int_{\nuHI}^{\nuHeI} e^{-\sigmaHI\langle\NHI\rangle}F_\nu^{\rm inc}/(h\nu)d\nu}{\int_{\nuHI}^{\nuHeI} F_\nu^{\rm inc}/(h\nu)d\nu}$. The shaded region marks the approximate transition between nebulae dominated by density-bound and ionization-bound channels. The figure shows that the turbulent $\HII$ regions allow high LyC escape fractions as the photons leak through narrow holes, but retaining high average $\HI$ column densities over the entire systems.}\label{fig:fesc_NHI}
\end{figure}

\subsection{Covering Fraction, Kinematics \& Spectral Hardness}

The leakage through the turbulent $\HII$ regions introduces the correlation of LyC escape fraction with the $\HI$ covering fraction, kinematics and spectral hardness. 

The probability distribution functions of the $\HI$ column densities in Figure~\ref{fig:NHI} show the two clear channels of LyC photons: one corresponding to the photoionized density-bound channels ($\NHI\approx10^{17-18}\rm~cm^{-2}$) where LyC escapes  and another corresponding to the neutral ionization-bound channels ($\NHI\approx10^{21-22}\rm~cm^{-2}$) where the I-fronts still reside within the system. As LyC photons escape through narrow photoionized channels, a large fraction of hydrogen can be retained in a neutral phase, allowing high LyC leakage with a high average $\HI$ column density $\langle\NHI\rangle$ (see Figure~\ref{fig:fesc_NHI}). As a result, $\langle\NHI\rangle$ of a system may not give a clear indicator of LyC leakage.

The quantity which is better correlated with LyC escape fractions is the $\HI$ covering fraction. We define the covering fraction, $f_{\rm cov}(>\NHI)$, as the fraction of slightlines with $\HI$ column densities greater than $\NHI$. We compute the covering fraction of optically thick sightlines with less than 1\% leakage at the Lyman limit\footnote{The Lyman limit column density $\NHI^{\rm LL}=1/\sigma_L^{-1}$ is the value that the gas starts to optically thick, but at the column density, the transmission $e^{-1}=0.37$ is still appreciably large.} corresponding to a $\HI$ column density more than $\NHI^{\rm thick}=-\sigma_L^{-1}\ln0.01\simeq7.3\times10^{17}\rm~cm^{-2}$. 

\begin{figure}
\hspace*{-0.5cm}
\includegraphics[width=1.1\columnwidth]{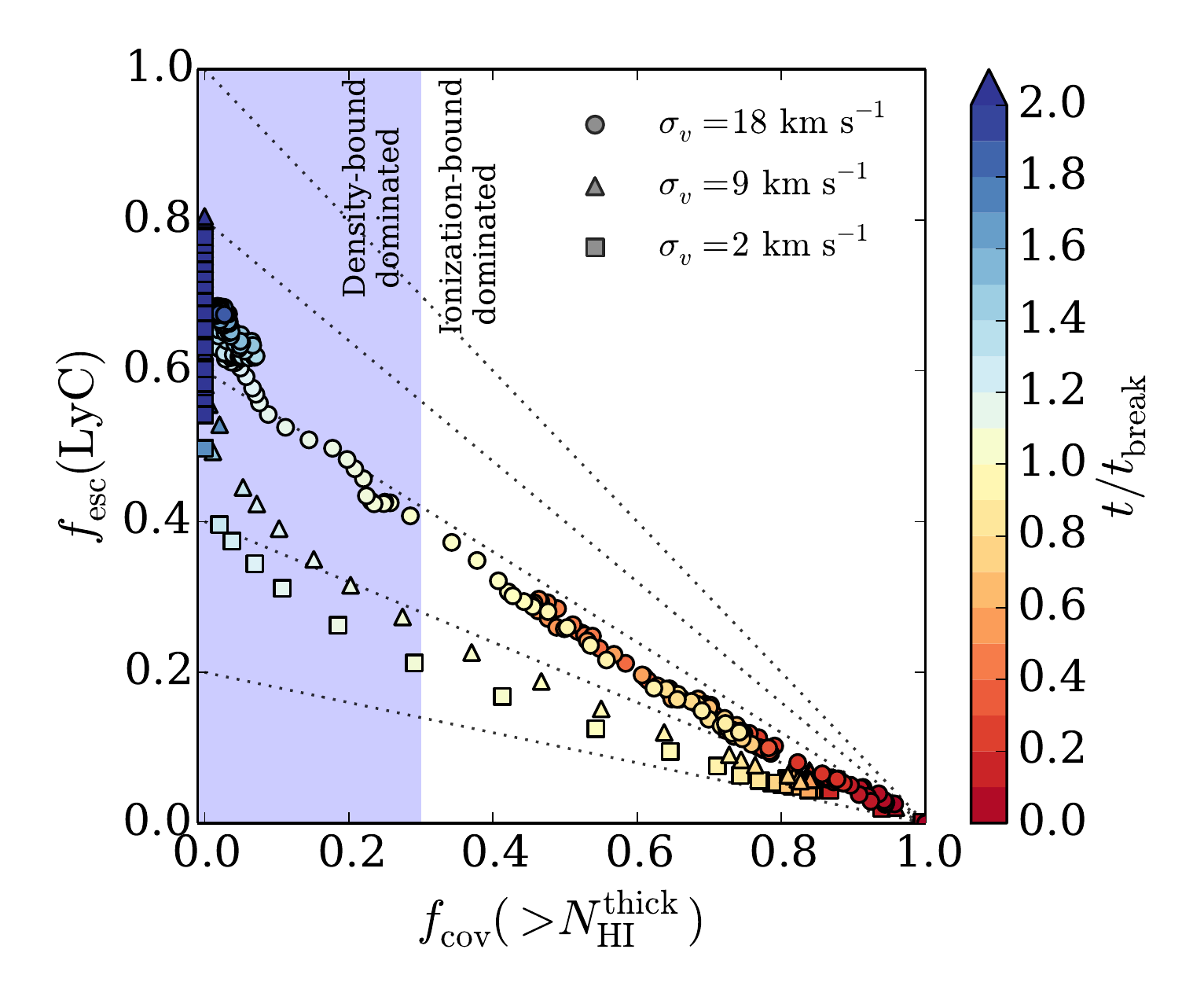}
\vspace{-0.8cm}
\caption{Correlation between LyC escape fraction and covering fraction of optically thick gas. The results from three runs (squares: \texttt{V2S\_f2e11\_RHD}, triangles: \texttt{V9S\_f2e11\_RHD}, circles: \texttt{V18S\_f2e11\_RHD}) are shown. The colors indicate the time normalized by the I-front breakout time of each simulation, $t/t_{\rm break}$. Linear relations, $f_{\rm esc}^{\rm LyC}\propto1-f_{\rm cov}$, with different slopes, $0.2,~0.4,~0.6,~0.8$ and $1.0$, are indicated by dotted lines. The shaded region marks the approximate transition between nebulae dominated by ionization-bound and density-bound channels before and after the breakout of average I-front.}\label{fig:fesc_fcov}
\end{figure}

Figure \ref{fig:fesc_fcov} shows the correlation between the LyC escape fraction and the covering fraction. Before the breakout of the average I-front, LyC escape is directly correlated with the $\HI$ covering fraction, which leads to a linear relation $f_{\rm esc}^{\rm LyC}\propto1-f_{\rm cov}(>\NHI^{\rm thick})$. However, as the photoionized channels are not empty, the escape fraction is lower because photons recombine inside them, causing $f_{\rm esc}^{\rm LyC}<1-f_{\rm cov}(>\NHI^{\rm thick})$.  While a low covering fraction is a necessary condition for a high LyC leakage, a measure of covering fraction places only an upper limit to the escape fraction \citep{Vasei16}.

The turbulent kinematics also influences the properties of the photoionized channels. For increasing turbulent velocities, there is a larger probability for lower densities to occur which leads to less recombination within the channels. This further implies a higher LyC escape faction at a given covering fraction. The resulting relation is therefore the combination of covering fraction and transmitted fraction of LyC photons through the photoionized channels, 
\begin{equation}
f_{\rm esc}^{\rm LyC}\approx f_{\rm tr}(\sigma_v)[1-f_{\rm cov}(>\NHI^{\rm thick})]. 
\end{equation}
In our simulations, we find the transmitted fractions are $f_{\rm tr}(\sigma_v)\approx\{0.28,~0.36,~0.55\}$ for $\sigma_v=\{2,~9,~18\}\rm~km~s^{-1}$ by fitting the linear relation to the numerical results.

After the average breakout of the I-fronts, because the system is dominated by density-bound channels, there is little correlation between the escape fraction and the covering fraction. The escape fraction is now regulated by the recombination rate in the $\HII$ region. At the density-bound dominated regime, the escape fraction is set by the balance between incident ionizing flux and the recombination rate in the photoionized medium,
\begin{equation}
f_{\rm esc}^{\rm LyC}\approx 1-\frac{\alpha_{\rm B}L}{F_{\rm ion}^{\rm inc}}\int n_e^2 P_V(n_e|\mathcal{M}_{\rm II})dn_e,\label{eq:22}
\end{equation}
where $P_V(n_e|\mathcal{M}_{\rm II})$ is the volume-weighted density probability distribution function of the ionized gas with a Mach number $\mathcal{M}_{\rm II}$. 

\begin{figure}
\centering
\vspace{0.2cm}
\includegraphics[width=0.99\columnwidth]{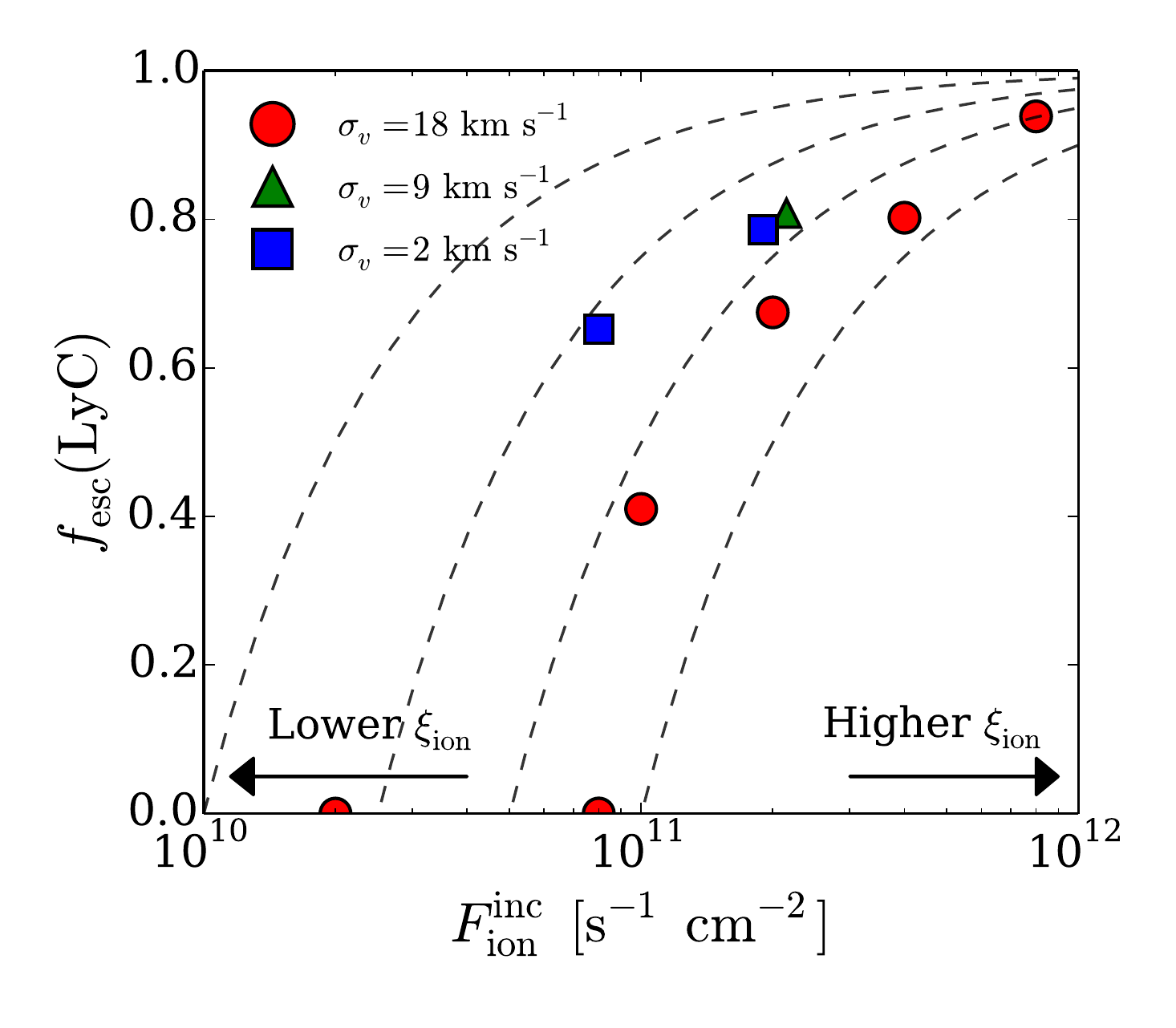}
\vspace{-0.4cm}
\caption{Correlation between LyC escape fraction and the incident ionizing flux $F^{\rm inc}_{\rm ion}$ (a proxy for spectral hardness $\xi_{\rm ion}$) at the end of simulations ($2~\rm Myr$). For all the points except for the first two red points, the average I-fronts have broken out already for all the simulations, therefore indicating the LyC escape fractions at the density-bound dominated limit. The first two red points are zero as the I-front does not break out before the end of simulation. The dashed lines indicate the expected scaling of $f_{\rm esc}^{\rm LyC}$ with $F^{\rm inc}_{\rm ion}$ in the density-bound leakage, $f_{\rm esc}^{\rm LyC}=1-({\rm const.})/F^{\rm inc}_{\rm ion}$. }\label{fig:spectral_hardness}
\end{figure}

In Figure~\ref{fig:spectral_hardness} we show the relation between the escape fraction and the incident ionizing flux.  At the density-bound dominated regime, the escape fraction increases with the incident ionizing flux (spectral hardness) as expected from Equation (\ref{eq:22}) (cf. Section~\ref{sec:theory}). The spectral hardness of the stellar population affects the LyC escape as harder sources induce more photoionization making photons easier to escape. This intrinsic dependence of LyC escape fraction on spectral hardness means that spectrally harder sources, common for higher redshifts and fainter objects \citep[e.g.][]{Matthee17,Harikane18}, may be able to deposit more ionizing photons into the surroundings because both the LyC escape fraction and ionizing photon production efficiency can increase the total escaping LyC luminosity, $\propto f_{\rm esc}^{\rm LyC}(\xi_{\rm ion})\xi_{\rm ion}\rm SFR$. 

\begin{figure*}
\centering
\includegraphics[width=1.5\columnwidth]{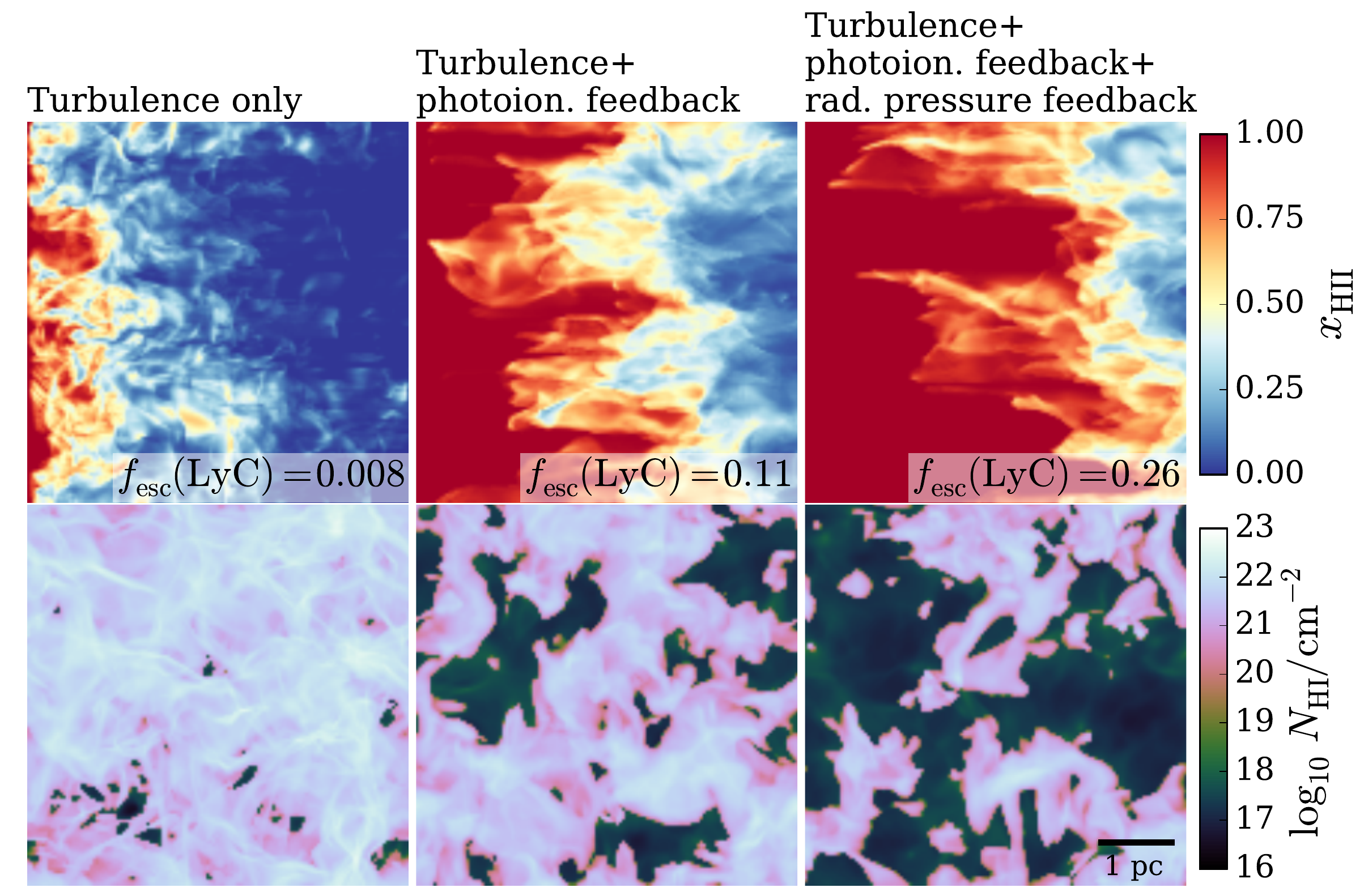}
\vspace{0.1cm}
\caption{Effect of turbulence and radiative feedback on LyC leakage. Projected maps of the mass-weighted ionized fractions and $\HI$ column densities in the three simulations (left: \texttt{V18S\_f2e11\_RT}, post-processed RT (turbulence only), middle: \texttt{V18S\_f2e11\_RHD-RP}, RHD without radiation pressure (turbulence$\,+\,$photoionization feedback), right: \texttt{V18S\_f2e11\_RHD}, full RHD (turbulence$\,+\,$photoionization feedback$\,+\,$radiation pressure feedback)) at 1 Myr. The LyC escape fractions are indicated at the bottom right corner of each panel.}\label{fig:map_rad_effect}
\end{figure*}

While the escape fraction can consistently be larger than $>10\%$ after the breakout, a higher level of turbulence will somewhat reduce the LyC escape in the density-bound dominated regime (see Figure~\ref{fig:spectral_hardness}). When the turbulent fluctuations is supersonic inside the $\HII$ region, the density fluctuations introduce the clumping of gas with an associated clumping factor,\footnote{The approximate equality is derived using the log-normal density probability distribution function for solenoidal turbulence (e.g. Federrath et al 2010). In our simulations, assuming the log-normal PDF for the ionized gas gives a clumping factor accurate to $\sim5\%$ to the simulated value.}
\begin{eqnarray}
\mathcal{C}(\mathcal{M}_{\rm II})=\frac{\langle n_e^2\rangle}{\bar{n}_e^2}&=&\int_{-\infty}^{\infty}\left(\frac{n_e}{\bar{n}_e}\right)^2P_V(n_e|\mathcal{M}_{\rm II})dn_e, \nonumber \\
&\approx&1+(\mathcal{M}_{\rm II}/3)^2.
\end{eqnarray}
This leads to a reduction of the escape fraction due to the enhanced recombination rate. For the $\sigma_v=18\rm~km~s^{-1}$ RHD turbulence simulation, We find $\mathcal{M}_{\rm II}\approx1.1$ and the clumping factor $\mathcal{C}(\mathcal{M}_{\rm II})\approx1.2$ in the photoionized gas. The effect becomes only prominent for a very high value of turbulent velocity dispersion $\sigma_v>12.8T_4^{1/2}\rm ~km~s^{-1}$ that can maintain supersonic fluctuations in the photoheated gas $\mathcal{M}_{\rm II}>1$. For a modest velocity dispersion $\sigma_v<12.8T_4^{1/2}\rm ~km~s^{-1}$, e.g. in the Milky-Way like GMCs, the photoionized gas remains subsonic $\mathcal{M}_{\rm II}<1$. Because the thermal gas pressure smooths out the density perturbations within a few sound crossing timescale faster than turbulent mixing, the density clumping is modest \citep{Konstandin12}. Indeed, in the simulations with $\sigma_v=2,~9\rm~km~s^{-1}$ ($\mathcal{M}_{\rm II}=0.1,~0.6$), the clumping factors of the ionized gas remain as $\mathcal{C}(\mathcal{M}_{\rm II})=1$. Subsonic turbulence inside the $\HII$ region thus has a negligible effect on the density-bound value of LyC escape fraction. 

In both the density- and ionization-bound dominated regimes, turbulent $\HII$ regions introduce a diversity in $f^{\rm LyC}_{\rm esc}$ for a given $\HI$ covering fraction and spectral hardness. 

\subsection{Role of Turbulence and Radiative Feedback}\label{sec:rad_effect}

The presence of turbulence alone is not a sufficient condition to trigger a high LyC leakage. The radiation-hydrodynamical coupling and the ability to ionize the gas beyond the classical Str\"omgren radius by the D-type I-front and the radiative feedback are important for regulating LyC leakage through a turbulent GMC. 

To illustrate this point, we compare a simulation in which the radiation-hydrodynamical coupling is switched off (i.e. post-processed RT) with the fiducial run with the same initial turbulence. As there is no dynamical response of gas by photoionization nor radiation pressure, the I-front remains as the R-type in the post-processed RT. In this case, if the initial gas column density is larger than,
\begin{equation}
N_{\rm H,0}>\frac{F_{\rm ion}^{\rm inc}}{\alpha_B\bar{n}_0}\approx1.15\times10^{21}T_4^{0.7}\left(\frac{\mathcal{U}}{10^{-2}}\right)\rm cm^{-2},\label{eq:trap}
\end{equation}
the I-front is kept trapped within a cloud. This is the regime similar to that studied by \citet{Safarzadeh16}. Comparison with the fiducial run is shown in Figure~\ref{fig:map_rad_effect}.  After several recombination times the R-type I-front reaches a steady state and is effectively frozen in.  Although the turbulence density fluctuations create lower column density channels by a few order of magnitudes around the mean \citep[e.g.][]{Federrath10}, there remains a substantial optical depth even along the lowest column density channels. Thus the total increase in LyC escape by turbulent fluctuations remains modest. For example, we find only $f_{\rm esc}^{\rm LyC}\simeq0.008$ in the R-type I-front simulation by post-processing RT whereas the full RHD simulation of the D-type I-front can reach $f_{\rm esc}^{\rm LyC}\gtrsim0.10$ after $\sim\,$Myr. The photoheating across the I-front and the associated shocks offer an effective means for LyC photons to evacuate efficiently through low column density channels.

The inefficiency of turbulent fluctuations when creating $\NHI<10^{17}\rm~cm^{-2}$ channels is not surprising as typical surface densities of GMCs are $\Sigma_{\rm cloud}\sim10-1000\rm~M_\odot~pc^{-2}$ \citep[][for a recent compilation and references therein]{Leroy15}, corresponding to the total hydrogen column density from the centre of the spherical cloud to the outer radius, $\NH=(3/4)(\Sigma_{\rm cloud}/m_{\rm H})\sim10^{21}-10^{23}\rm~cm^{-2}$. This means that about $4-7$ orders of magnitude fluctuations in column densities within a cloud are required to produce $\NHI\lesssim10^{17}\rm~cm^{-2}$ channels. In fact, following \citet{Brunt10b,Brunt10a} and \citet{Thompson&Krumholz16}, we can estimate the fraction of the I-fronts being trapped in a turbulence without radiative feedback as
\begin{equation}
f_{\rm trap}=\int^{\infty}_{s_{\rm trap}}P_{\rm H{\scriptscriptstyle\,I}}(s|\mathcal{M}_{\rm I})ds =\frac{1}{2}{\rm erfc}\left[\frac{\sigma^2_{\ln\NH}+2s_{\rm trap}}{\sqrt{8\sigma^2_{\ln\NH}}}\right],
\end{equation}
where $P_{\rm H{\scriptscriptstyle\,I}}(s|\mathcal{M}_{\rm I})$ is the log-normal probability distribution of column density and $s_{\rm trap}=\ln(\NH^{\rm trap}/\bar{\NH})$ with $\NH^{\rm trap}\approx1.15\times10^{21}\rm~cm^{-2}$ (see Equation (\ref{eq:trap})). The standard deviation is given by
\begin{equation}
\sigma^2_{\ln\NH}\approx\ln\left[1+R(\mathcal{M}_{\rm I}/3)^2)\right],
\end{equation}
and $R=\frac{1}{2}\left(\frac{3-\alpha}{2-\alpha}\right)\left(\frac{1-\mathcal{M}_{\rm I}^{2(2-\alpha)}}{1-\mathcal{M}_{\rm I}^{2(3-\alpha)}}\right)$ with $\alpha=2.5$ being the power-law index of density power spectrum $P(k)\propto k^{-\alpha}$ \citep[e.g.][for a review]{Krumholz14}. The estimate suggests only $1-f_{\rm trap}\approx0.06\%$ of I-fronts in the $\mathcal{M}_{\rm I}\approx10$ gas in a $\bar{\NH}\approx10^{22}\rm~cm^{-2}$ cloud can exit the system. For $\mathcal{M}_{\rm I}\approx20$, the fraction is still $1-f_{\rm trap}\approx1\%$. Although a higher Mach turbulence and the associated increase in intermittency would open up more low column density channels \citep{Hopkins13,Federrath13}, substantial LyC escape still require a radiative feedback in addition to turbulence (cf. Figure~\ref{fig:map_rad_effect}).

Among the two radiative feedback mechanisms -- photoionization heating and radiation pressure -- the direct ionizing radiation pressure plays a secondary role in evacuating the gas through the low column density channels, in agreement with previous studies \citep{Rosdahl15}. The relative importance of photoionization and radiation pressure is easy to understand by taking the thermal-to-radiation pressure ratio \citep{Lopez14,McLeod18},
\begin{align}
\frac{P_{\rm th}}{P_{\rm rad}}&=\frac{2k_Bn_eT}{\langle h\nu\rangle F_{\rm ion}^{\rm inc}/c} \\
&\approx 3.8T_4\left(\frac{ n_e}{200\rm~cm^{-3}}\right)\left(\frac{F_{\rm ion}^{\rm inc}}{2\times10^{11}\rm~s^{-1}~cm^{-2}}\right)^{-1}. \nonumber  
\end{align}
The impact of ionizing radiation pressure less effective than the thermal pressure by photoionization heating even in the plane-parallel geometry, which will be reduced further in a spherical geometry by the geometric factor of $r^{-2}$. Ly$\alpha$ radiation pressure may be important although the exact degree of the impact is still unclear \citep{Dijkstra08,Dijkstra09,Smith17,Kimm18}. In the regimes studied, the shocks induced by photoionization heating and turbulent fluctuations are likely the major modes of regulating the opening of low column density channels. In summary, we find that turbulence is not a sole agent to regulate LyC leakage in the $\HII$ regions. Radiative feedback {\it and} turbulence are what regulate the LyC leakage, in particular, the photoheated I-front shock provides an effective means for evacuating the gas through the openings of turbulence-generated channels.

\section{Ly$\boldsymbol\alpha$-LyC Connection}\label{sec:Lya-LyC}

\subsection{Physics of Ly$\alpha$ Line Formation}

\begin{figure}
\centering
\includegraphics[width=\columnwidth]{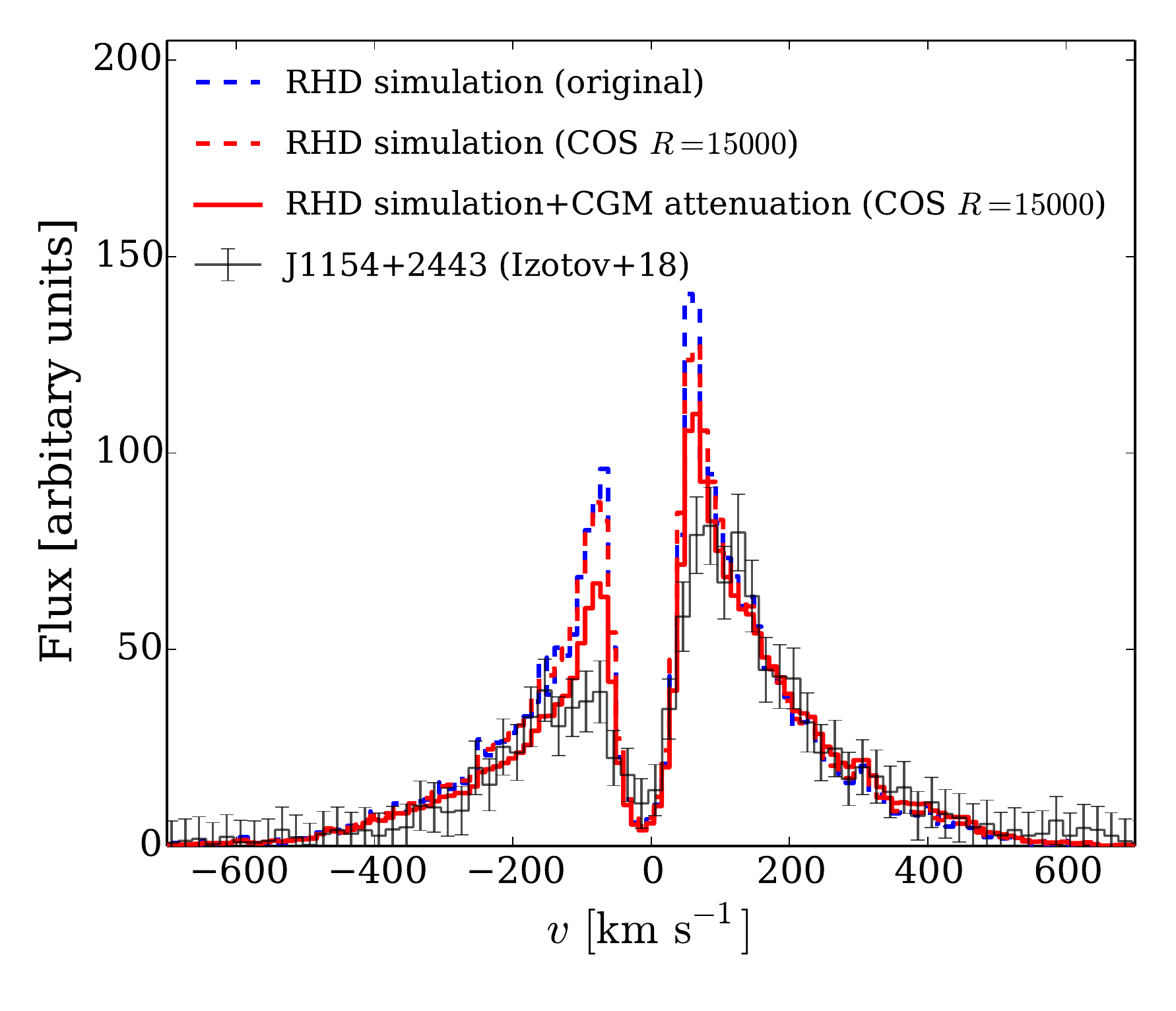}
\vspace{-0.8cm}
\caption{Comparison of the emergent Ly$\alpha$ line profile from the RHD turbulence simulation (\texttt{V18S\_f2e11\_RHD} at $t=0.5\rm~Myr$) with the observed COS spectrum of a LyC-leaking galaxy, J1154+2443 \citep{Izotov2018a}. The three simulated spectra correspond to one at the spectral resolution of the simulation (blue dashed), one at the COS resolution ($R=15000$) (red dashed), and one with the CGM attenuation model of \citet{Kakiichi&Dijkstra18} at the COS resolution (red solid). Regardless of the additional uncertainty from the CGM, Ly$\alpha$ transfer through the turbulent LyC-leaking $\HII$ region can explain the observed spectrum reasonably well.}\label{fig:Lya}
\end{figure}

\begin{figure*}
\centering
\includegraphics[width=\textwidth]{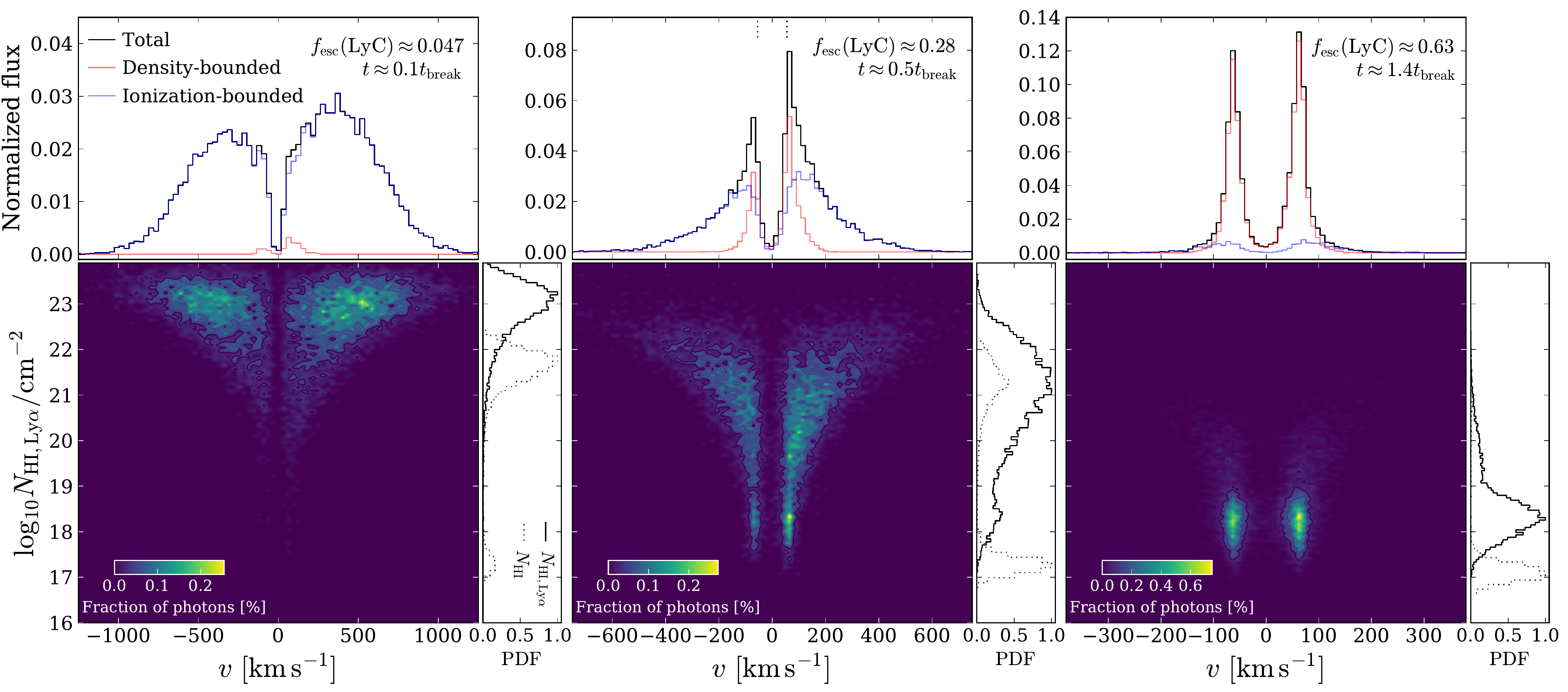}
\caption{Two-dimensional Ly$\alpha$ spectra decomposed with respect to the total $\HI$ column density integrated along the path of a Ly$\alpha$ photons, $\NHILya$. The projection along each axis corresponds to the total Ly$\alpha$ profile (black, top inset) and the $\NHILya$ probability distribution function (solid, right inset) normalized at their maximum values. The red and blue lines show the contributions of the photons experienced $\log_{10}\NHILya/{\rm cm^{-2}}<20$ and $>20$, which approximately indicates the photons through photoionized LyC escape channels and those though optically thick density-bound channels. For reference, a dotted line shows the physical $\NHI$-distribution for each snapshot. The spectra are drawn from \texttt{V18S\_f2e11\_RHD} simulation at the three different times bracketing the three representative cases (ionization-bound dominated limit, a mixed case, and density-bound dominated limit).}\label{fig:spec2d}
\end{figure*}

The Ly$\alpha$ line profiles emerging from the RHD turbulence depend upon to the properties of LyC leakage. The detail of Ly$\alpha$ transfer is described elsewhere \citep[see e.g.][]{Dijkstra17}. Here we summarise the relevant aspects of Ly$\alpha$ line formation for LyC-leaking turbulent $\HII$ regions.

The LyC leakage through a RHD turbulence produces the two distinctive passages for Ly$\alpha$ photons; one with low column densities $\NHI\approx10^{17-18}\rm~cm^{-2}$ allowing LyC escape (density-bound channels) and another with $\NHI\approx10^{21-22}\rm~cm^{-2}$ that remains optically thick to LyC (ionization-bound channels) (cf. Figure~\ref{fig:NHI}). This bimodal $\HI$ distribution corresponds to that of the `picket-fence' or `Holes' model often used to interpret observations \citep{Zackrisson13,Reddy16,Steidel18}. Here, such geometry of $\HI$ gas arises naturally as a consequence of the RHD turbulence. 

The Ly$\alpha$ optical depth at a frequency, $x=(\nu-\nu_\alpha)/\Delta\nu_D$, is $\tau_\alpha=\tau_{\alpha,0}\phi(x)$ where $\nu_\alpha$ is the resonant frequency, $\Delta\nu_D=\nu_\alpha b/c$ is the Doppler width ($b$ is the Doppler $b$-parameter), $\phi(x)$ is the Voigt profile, $\tau_{\alpha,0}=\sigma_{\alpha,0}\NHI$ is the optical depth at line center, and $\sigma_{\alpha0}\approx5.9\times10^{-14}T_4^{-1/2}\rm~cm^{2}$ is the line centre Ly$\alpha$ cross section. The gas above the $\HI$ column density,
\begin{equation}
\NHI>1/\sigma_{\alpha,0}\approx1.7\times10^{13}T_4^{1/2}\rm~cm^{-2},
\end{equation}
is optically thick to the photons emitted at line center. Therefore, Ly$\alpha$ photons are more susceptible to the amount of (residual) neutral hydrogen compared to the column density ($\NHI\lesssim10^{17-18}\rm~cm^{-2}$) required for LyC photons to escape.
For example, in the fiducial simulation, the photoionized LyC escaping channels (having the residual $\HI$ fractions, $\xHI\approx\alpha_{\rm B} n_{\rm H}/\Gamma\sim10^{-4}-10^{-5}$, and the gas density, $n_{\rm H}\sim200\rm~cm^{-3}$) have an average $\HI$ column density of
\begin{equation}
\bar{N}_{\mbox{\tiny H\Rmnum{1}},\rm channel}\approx 2.4\times10^{17} \rm~cm^{-2},
\end{equation}
filling up $1-f_{\rm cov}$ fraction of sightlines around ionizing sources. The Ly$\alpha$ photons emitted at the line center therefore cannot freely steam out. Instead, they experience appreciable scattering events before escape. \textit{The resulting Ly$\alpha$ profile is therefore strongly influenced by the availability, the {H\,{\scriptsize \it I}} column density, and the kinematics of LyC escape channels.}

The turbulence also introduces both density and velocity fluctuations which would influence the fate of the Ly$\alpha$ photons. \citet{Gronke16,Gronke17} detailed the Ly$\alpha$ transfer mechanism through a clumpy medium and found that Ly$\alpha$ photons can escape either ({\it i}) via a `single flight' or `excursion' after core or wing scatterings, i.e. similar to through an homogeneous slab \citep{Osterbrock1962,Adams72}, or ({\it ii}) via `random walk' between clumps \citep{Hansen&Oh06}. These two different modes of Ly$\alpha$ escape subsequently leave an imprint on the emergent Ly$\alpha$ spectrum, most easily identified by the flux at line center. Since the filling factor of the optically thick gas to Ly$\alpha$ photons is high both inside the $\HII$ region and in the underdense regions of turbulence, there is little room for Ly$\alpha$ photons to freely travel between the clumps in a turbulent $\HII$ region. therefore, for our parameter space studied, the photons escape primarily by the former mechanism, i.e. via single flight or excursion.\footnote{Although we find Ly$\alpha$ escape mostly via single flight or excursion, we expect the contribution from escape via random walk may increase if an extreme turbulence maintains a high level of density fluctuations in the $\HII$ region. Ly$\alpha$ escape via random walk through clumpy channels will then contribute to the residual non-zero flux at line center.} If the $\HI$ column density of a channel is optically thick to the Doppler core, but optically thin to the Lorentzian wing, Ly$\alpha$ photons escape via a single flight after the frequency is shifted out of the core. Such escape is dominant in low-$\NHI$ channels with \citep{Osterbrock1962,Zheng&Miralda-Escude02}
\begin{equation}
\NHI\lesssim\frac{1}{\sigma_{\alpha,0}\phi(x_*)}\approx7\times10^{17}\rm~cm^{-2},
\begin{tabular}{c}
\small escape via single flight \\[-0.1cm]
\small after core scatterings
\end{tabular}
\end{equation}
where $x_*=3.26$ is the core-wing transition frequency at $T=10^4\rm~K$. On the other hand, if the gas remains optically thick far in the wing, Ly$\alpha$ photons primarily escape via diffusion in the frequency space after multiple wing scatterings, that is, via excursion. This escape is dominant in high-$\NHI$ channels with \citep{Neufeld90,Dijkstra06}
\begin{equation}
\NHI\gtrsim\frac{10^3}{\sigma_{\alpha,0}a_v}\approx4\times10^{19}\rm~cm^{-2},
\begin{tabular}{c}
\small escape via excursion \\[-0.1cm]
\small after wing scatterings
\end{tabular}
\end{equation}
where $a_v=4.7\times10^{-4}$ is the Voigt parameter at $T=10^4\rm~K$. The resulting Ly$\alpha$ spectrum in the RHD turbulence is therefore controlled by the combination of the above mechanisms.

\begin{figure*}
\includegraphics[width=1.21\columnwidth]{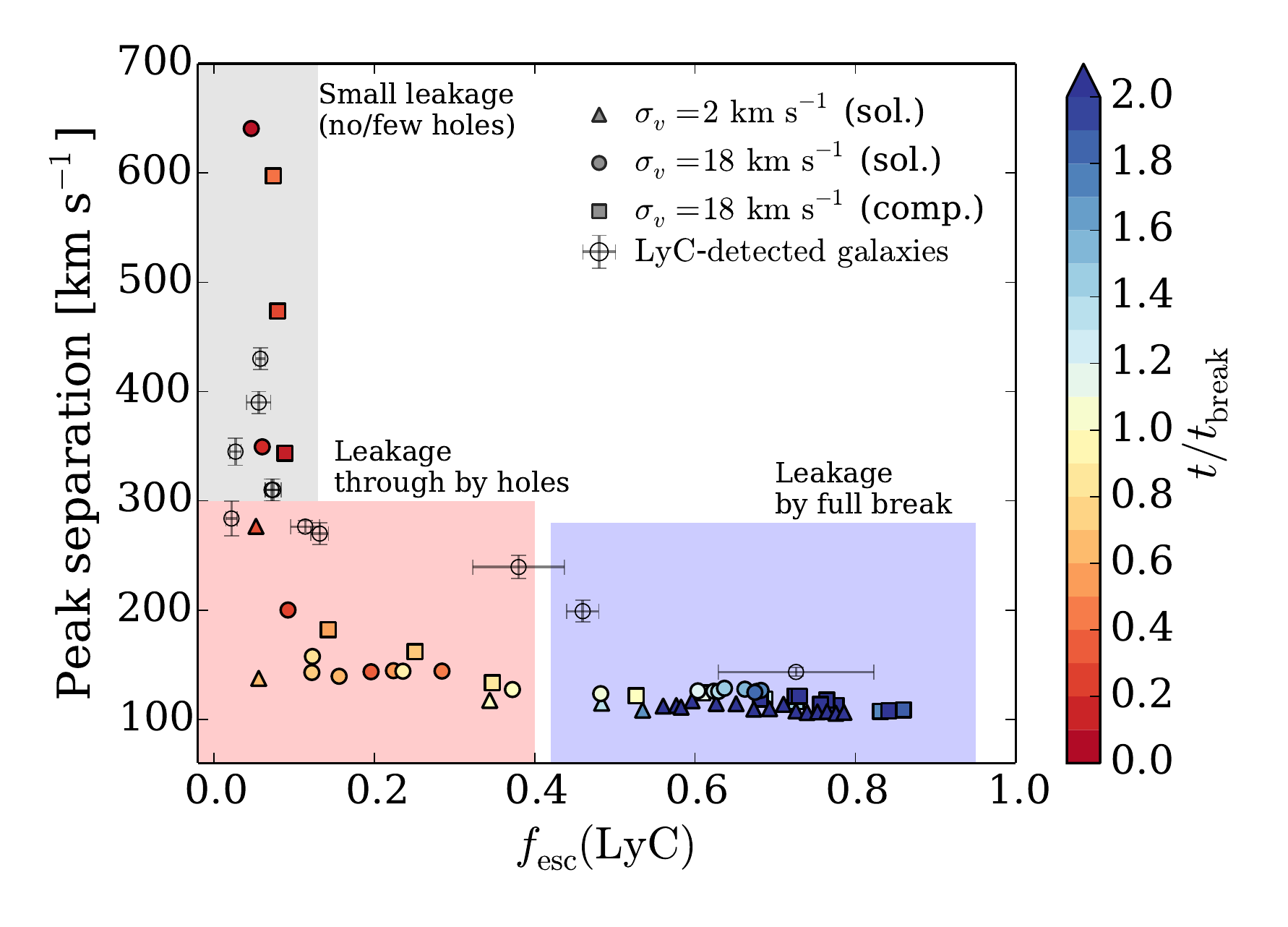}
\hspace{-2cm}
\includegraphics[width=1.13\columnwidth]{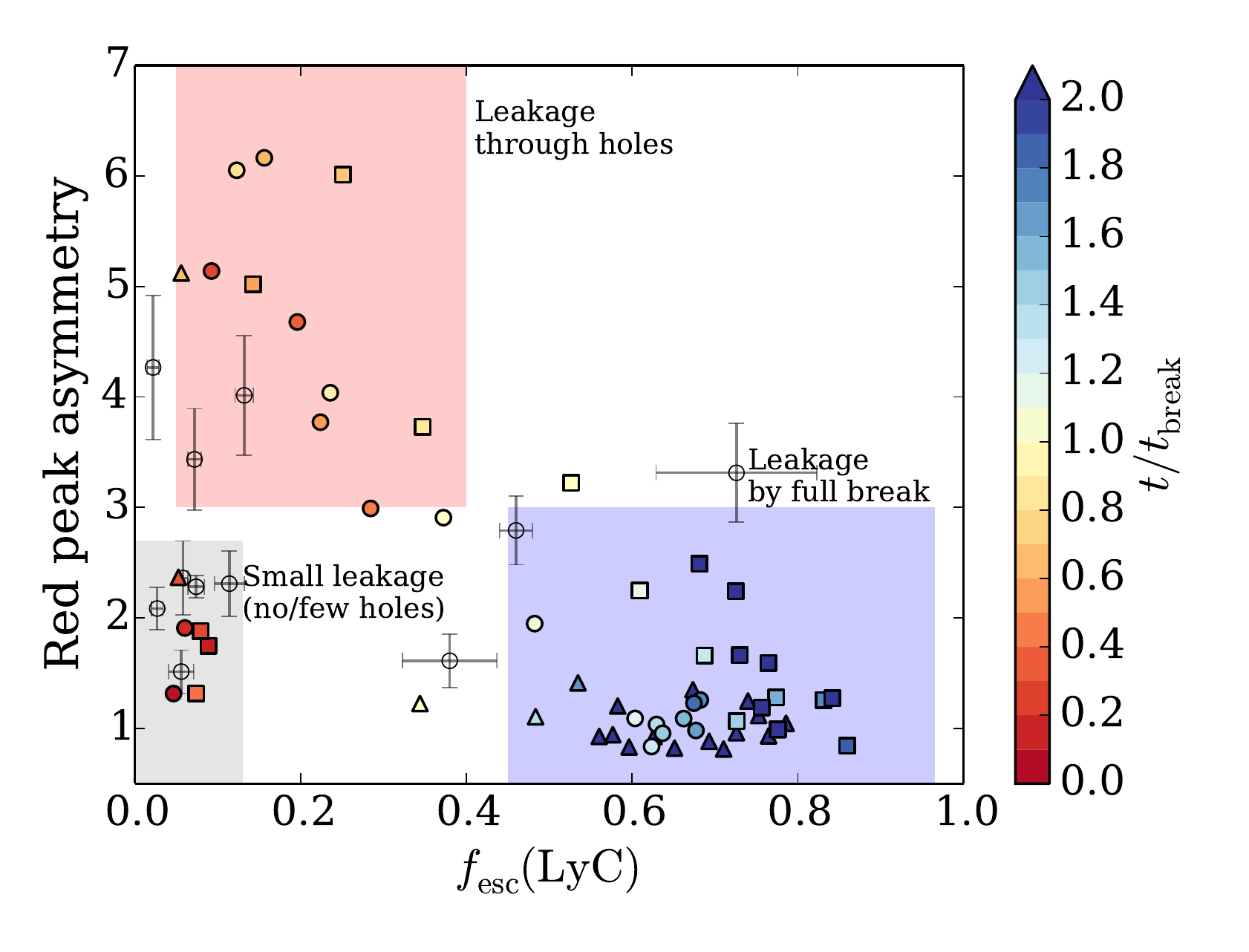}
\vspace{-0.8cm}
\caption{({\it Left}): Relation between the Ly$\alpha$ peak separation $\Delta v_{\rm peak}$ and LyC escape fraction $f_{\rm esc}^{\rm LyC}$. ({\it Right}): Relation between the Ly$\alpha$ red asymmetry $A_f$ and LyC escape fraction $f_{\rm esc}^{\rm LyC}$. The colors indicate the time normalized by the I-front breakout time of each simulation, $t/t_{\rm break}$, and the different symbols correspond to the three different runs (triangles: \texttt{V2S\_f2e11\_RHD}, circles: \texttt{V18S\_f2e11\_RHD}, squares: \texttt{V18C\_f2e11\_RHD}). The errorbars indicate the observed $z\sim3$ LyC-detected galaxies from \citep{Izotov2016,Izotov2018a,Izotov2018b}. The shaded regions are marked to guide the different regimes of LyC leakage and the associated values of the peak separation and asymmetry parameter (see text).}\label{fig:fesc-lya}
\end{figure*}

In summary, the above Ly$\alpha$ transfer mechanism in a turbulent $\HII$ region produces diverse Ly$\alpha$ line profiles, including narrow double peak profiles with high LyC escape fractions. In Figure~\ref{fig:Lya} we show a case that reproduces the observed Ly$\alpha$ profile of a $z\sim0.3$ LyC-leaking galaxy \citep[J1154+2443,][]{Izotov2018a}. Although we admittedly chose a simulation that resembles the observation, the match is worth noting given that the required multiphase structure and the subsequent Ly$\alpha$ transfer naturally emerge from a RHD turbulence simulation representing the $\HII$ region in a GMC.    

To understand the formation mechanism of Ly$\alpha$ spectra in detail, in Figure~\ref{fig:spec2d} we have decomposed three representative Ly$\alpha$ spectra as a function of the integrated $\HI$ column density seen by each Ly$\alpha$ photon. $\NHILya$ is the $\HI$ column density integrated along a path of a Ly$\alpha$ photon. This allows us to quantify the contributions of Ly$\alpha$ photons escaped through various paths to the total Ly$\alpha$ profiles (top insets). Because a Ly$\alpha$ photon travels in a zigzag path through a medium by scatterings, its integrated path is longer than the length of the simulation box, leading $\NHILya$ to be generally larger than the physical $\NHI$. Figure~\ref{fig:spec2d} clearly shows that the two distinct passages of Ly$\alpha$ photons through density- and ionization-bound channels contribute differently to the various components (peaks and broad wings) of the Ly$\alpha$ profile. The origin of each component is detailed below.

\subsection{Origin of the Peak Separation}

In a LyC-leaking $\HII$ region, the Ly$\alpha$ peak separation is determined by \textit{the {H\,{\scriptsize \it I}} column density and temperature of the photoionized LyC escaping channels}. Figure~\ref{fig:spec2d} shows that when a system shows a high LyC leakage $f_{\rm esc}^{\rm LyC}\gtrsim10\%$, the Ly$\alpha$ photons propagating through density-bound channels dominate the location of the Ly$\alpha$ peaks, whereas when there are no or few holes through which LyC photon can escape, the broad peak separation is produced. In the latter case, as the majority of the I-fronts are still bound within a cloud, the most of Ly$\alpha$ photons need to escape by scattering through optically thick, ionization-bound channels of $\NHI>10^{21}\rm~cm^{-2}$. This inevitably leads to a broad Ly$\alpha$ peak separation and low LyC escape fraction. On the other hand, the formation of a narrow peak separation occurs as soon as the LyC escape channels open up, and Ly$\alpha$ photons can escape through paths of lower column density \citep{Dijkstra16a,Eide18}. As this can happen before the breakout of the average I-fronts $t<t_{\rm break}$, the separation remains approximately constant even after the breakout ($\Delta v_{\rm peak}\approx 100-200\rm~km~s^{-1}$). This reflects the fact that the peak separation is controlled by the $\HI$ column density of the photoionized channels, but not by the total averaged $\HI$ of the entire medium $\langle\NHI\rangle$ (cf. Figure~\ref{fig:fesc_NHI}). 

The resulting correlation between the peak separation and LyC escape fraction is shown in Figure~\ref{fig:fesc-lya}. The simulations show the anti-correlation in agreement with the observed trend \citep{Verhamme17,Izotov2018b}. The three representative Ly$\alpha$ spectra (see Figure~\ref{fig:spec2d}) occupy different regions of the diagram, illustrating how different LyC leakage mechanisms can lead to the Ly$\alpha$ peak separation - LyC escape fraction correlation.

The peak separation of a LyC-leaking $\HII$ region can be estimated analytically. Because at the $\HI$ column density of the density-bound channels $\NHI\approx10^{17-18}\rm~cm^{-2}$, the gas is optically thin to the wing, Ly$\alpha$ photons escape freely once the frequency is shifted out of the Doppler core (i.e. via single flight). Therefore by solving $\tau_\alpha=\tau_{\alpha,0}e^{-x^2}<1$ we find the escape frequency of $x>\sqrt{\ln\tau_{\alpha,0}}$ \citep{Osterbrock1962,Zheng&Miralda-Escude02}. The peak separation is then estimated by 
\begin{align}
    \Delta v_{\rm peak}&=2c_{\rm s,II}\sqrt{\ln\sigma_{\alpha,0}\bar{N}_{\mbox{\tiny H\Rmnum{1}},\rm channel}}, \nonumber \\
    &\simeq25.6T_4^{1/2}\sqrt{\ln\left(\frac{\bar{N}_{\mbox{\tiny H\Rmnum{1}},\rm channel}}{1.7\times10^{13}T_4^{1/2}\rm~cm^{-2}}\right)}.
\end{align}
Evaluating at the values found in the simulation $\bar{N}_{\mbox{\tiny H\Rmnum{1}},\rm channel}=2.4\times10^{17}\rm~cm^{-2}$ and $T=2\times10^4\rm~K$ we find 
\begin{equation}
\Delta v_{\rm peak}\approx110\rm~km~s^{-1}
\end{equation} 
in agreement with the simulated peak separation. Note that, in this case, the peak separation depends weakly on the $\HI$ column density of the channels ranging only $\sim90-120\rm~km~s^{-1}$ over $16<\log_{10}\NHI/{\rm cm^{-2}}<18$, but more sensitively on the temperature of the channels to nearly $\propto T^{1/2}$ (more precisely, the Doppler $b$-parameter). The peak separation can vary from $\sim80$ to $235\rm~km~s^{-1}$ over $T=10^{4-5}\rm~K$ ($b=12.8-40.5\rm~km~s^{-1}$). This is a direct consequence of the Doppler core scattering. As a corollary, it is possible that the leakage of hot gas through the channels in the $\HII$ regions \citep{Lopez11,Lopez14} to elevate the Ly$\alpha$ peak separation. This contrasts with the estimate $\Delta v_{\rm peak}\simeq300T_4^{1/6}(\NHI/10^{20}{\rm cm^{-2}})^{1/3}\rm~km~s^{-1}$ for $\NHI\gtrsim10^{18}\rm~cm^{-2}$ \citep[e.g.][]{Adams72,Dijkstra17} where the peak separation depends mostly on the $\HI$ column density. This regime is only valid when Ly$\alpha$ photons escape via excursion, thus only applies to an ionization-bound dominated system with high $\HI$ coverage in nearly all directions. This dichotomy is in accordance with the Monte-Carlo calculation of \citet{Verhamme15} where the peak separation of a homogeneous shell spans these two regimes depending on the $\HI$ column densities. 

The fact that the Ly$\alpha$ peak separation of a high LyC leaking medium reflects the $\HI$ column density of the escape channels instead of the average $\HI$ of the system has an observational implication. 21-cm observation reveals abundant $\HI$ gas of mass $M_{\rm HI}\sim10^{7-9}\rm~M_\sun$ in blue compact dwarf galaxies \citep{Thuan16,McKinney19} and in the LARS sample selected to be comparable to high-$z$ LAEs and LBGs \citep{Pardy14,Pardy16,Puschnig17}. This corresponds to the average $\HI$ column density of $\langle\NHI\rangle\sim5\times10^{19}{\rm~cm^{-2}}(M_{\rm HI}/10^7{\rm~M_\sun})(R/5\rm~kpc)^{-2}$. At face value this seems to exceed the value that allows narrow Ly$\alpha$ peak separation and high LyC leakage. This, however, can be explained if a galaxy consists of an ensemble of LyC-leaking $\HII$ regions that are multiphase, for example, those generated by the RHD turbulence. In such a system, while the Ly$\alpha$ peak separation and LyC leakage are regulated by escape of the photons through low column density channels, the majority of $\HI$ still resides in the neutral phase, keeping the average $\HI$ column density of the system high as observed by the 21-cm line. Similar argument may apply for the high $\HI$ coverage found in GRB-host galaxies \citep{Tanvir19}. However as both the covering fraction and the derived column density depends on the modelling of the absorption lines, we defer the detailed comparison to future work.

\subsection{Origin of the Peak Asymmetry}
\label{sec:peak_shape}

The Ly$\alpha$ peak asymmetry also reflects the multiphase nature of the turbulent $\HII$ regions. When Ly$\alpha$  escapes, some of Ly$\alpha$ photons need to undergo multiple scattering events through optically thick channels before the complete I-front breakout occurs. These photons diffuse more in frequency space and produces a broad wing component ($|v|\gtrsim250\rm~km~s^{-1}$) to the emergent Ly$\alpha$ profile. This component is clearly seen in Figure~\ref{fig:spec2d} when the optically thick ionization-bound channels exist. 

The presence of multiple routes of Ly$\alpha$ escape can be quantified by the asymmetry parameter of the red Ly$\alpha$ peak, $A_f$, defined as the ratio of the blue-to-red flux of the red peak,
\begin{equation}
A_f=\frac{\int_{\lambda_{\rm peak}^{\rm red}}^\infty f_\lambda d\lambda}{\int_{\lambda_{\rm valley}}^{\lambda_{\rm peak}^{\rm red}} f_\lambda d\lambda}
\end{equation}
where $f_\lambda$ is the flux, $\lambda_{\rm peak}^{\rm red}$ is the wavelength at the red peak and $\lambda_{\rm valley}$ is the wavelength at the minimum between red and blue peaks. This is similar to the asymmetry statistics introduced by \citet{Rhoads03}. Figure~\ref{fig:fesc-lya} shows the relation between the red peak asymmetry parameter and LyC escape fraction. The shaded regions in the diagram mark the approximate regions occupied by the different LyC leakage mechanisms (associated with the three representative Ly$\alpha$ spectra shown in Figure~\ref{fig:spec2d}). The simulations indicate that the Ly$\alpha$ peak asymmetry is high ($A_f\gtrsim3$) when both optically thin and thick channels co-exist whereas the asymmetry is low ($A_f\lesssim3$) when the medium is dominated either by ionization-bound channels or density-bound channels, in which only one type of Ly$\alpha$ escape is possible (either via single flight or excursion). This means that the anisotropic LyC leakage through holes in a turbulent $\HII$ region and isotropic LyC leakage from a fully density-bound $\HII$ region can be distinguishable by the measurement of the peak asymmetry. While the both mechanisms allow a high LyC escape fraction $f_{\rm esc}^{\rm LyC}\gtrsim10~\%$, the former favors a high asymmetry parameter ($A_f\gtrsim3$) whereas the latter is associated with a low asymmetry parameter ($A_f\lesssim3$). 

We have measured the red peak asymmetry parameter using the archival COS Ly$\alpha$ spectra of the $z\sim0.3$ LyC-detected sample of \citet{Izotov2016,Izotov2018a,Izotov2018b} (see Appendix B). Comparison with the simulation suggests that there may be a tentative trend indicating various LyC leakage mechanisms in the observed LyC-detected galaxies. The detailed analysis of the individual objects with the synthetic $\HII$ regions is needed to confirm the trend. 

\subsection{Role of Outflow and Turbulence on Ly$\alpha$ Line}

\begin{figure}
\centering
\vspace{0.2cm}
\includegraphics[width=0.9\columnwidth]{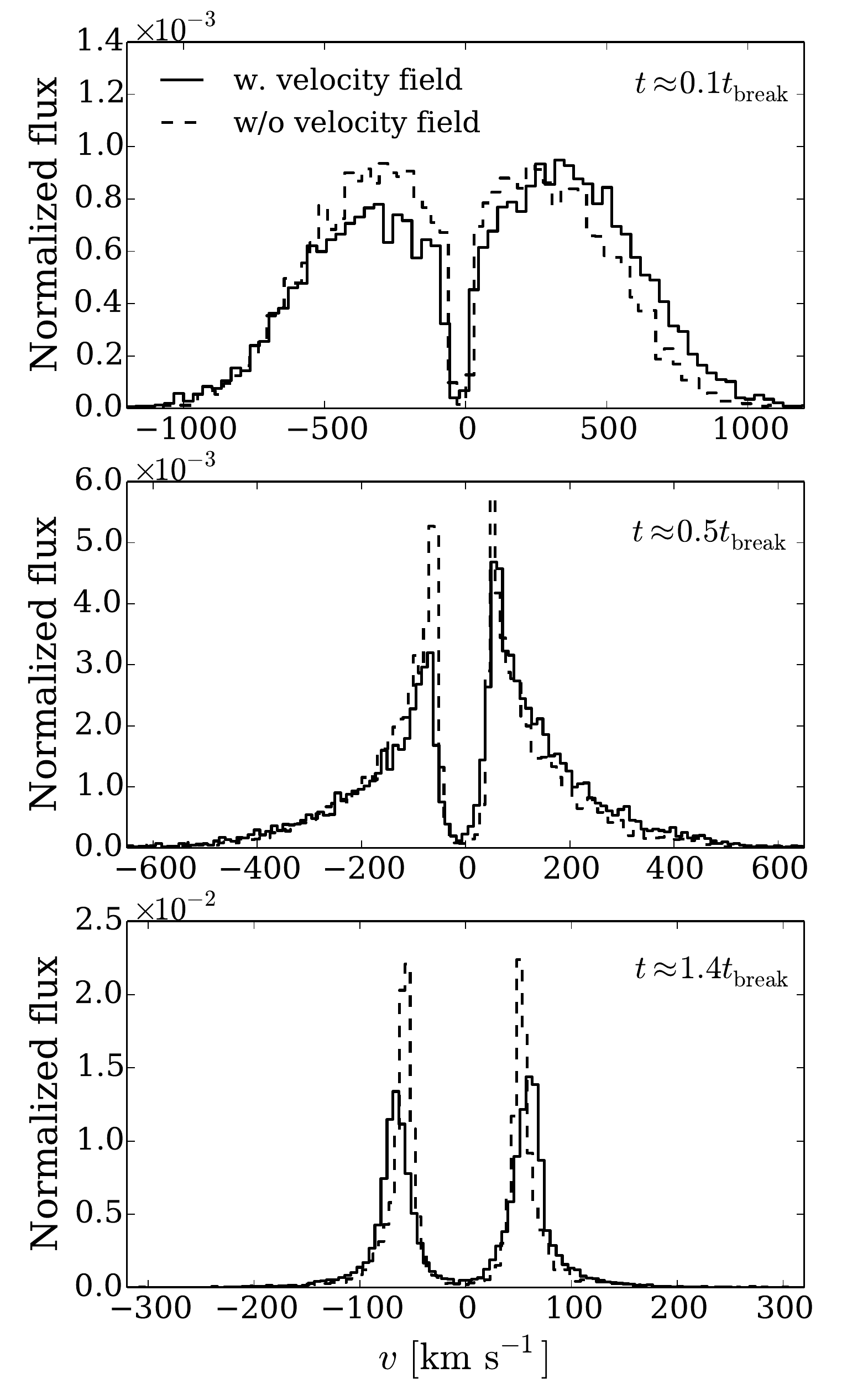}
\vspace{-0.2cm}
\caption{The influence of the gas kinematics of the $\HII$ region on Ly$\alpha$ line profile. The Ly$\alpha$ spectra from the Monte-Carlo Ly$\alpha$ simulation based on \texttt{V18S\_f2e11\_RHD} including velocity field (solid) and without velocity field (dashed) are shown for the three representative snapshots.}\label{fig:vel_effect}
\end{figure}

In order to examine the effect of kinematics on the Ly$\alpha$ line profile, Figure \ref{fig:vel_effect} compares the two sets of the Monte-Carlo Ly$\alpha$ RT simulations with and without the velocity fields from the RHD turbulence. 

The outflow driven by the photoionization heating (and radiation pressure) produces the enhancement of the red peak relative to the blue peak. The effect of outflow on the line profile is well known \citep[e.g.][]{Dijkstra06,Verhamme06}. Here, the enhancement is modest as the outflow velocity $\langle v\rangle_{\rm outflow}$ is only a few tens of $\rm~km~s^{-1}$ corresponding to approximately the expansion velocity of the I-front. Note that our RHD turbulence simulations show a double-peaked profile because the outflow velocity of the photoionized channel is 
\begin{equation}
    \langle v\rangle_{\rm outflow}<b\sqrt{\ln\sigma_{\alpha,0}\bar{N}_{\mbox{\tiny H\Rmnum{1}},\rm channel}},
\end{equation} 
where $b\sqrt{\ln\sigma_{\alpha,0}\bar{N}_{\mbox{\tiny H\Rmnum{1}},\rm channel}}\approx40T_4^{1/2}\rm km~s^{-1}$ at $\bar{N}_{\mbox{\tiny H\Rmnum{1}},\rm channel}=2.4\times10^{17}\rm~cm^{-2}$ ($b\approx c_{\rm s,II}$, see below).
At this modest outflow velocity, the gas remains optically thick to the Ly$\alpha$ photons emitted at line center. Therefore the photons will be absorbed and undergo core scatterings before escape. On the other hand, if the outflow velocity becomes faster such that $\langle v\rangle_{\rm outflow}>b\sqrt{\ln\sigma_{\alpha,0}\bar{N}_{\mbox{\tiny H\Rmnum{1}},\rm channel}}$, the photoionized channels are no longer optical thick to the photons emitted at line center, which now can freely escape without any interaction. Therefore, we expect if the photoionized gas is accelerated further by other (stellar) feedback the emergent Ly$\alpha$ profile will show a single peak component with a large Ly$\alpha$ flux at line center.

In Figure \ref{fig:vel_effect} the effect of turbulent velocity on the peak separation appears small. This is somewhat surprising as the na\"ive inclusion of the turbulence via a simple rescaling of the Dopper $b$-parameter, $\Delta v_{\rm peak}\propto b={c_{\rm s,II}}\sqrt{1+\mathcal{M}_{\rm II}^2}$ suggests nearly $\sim50\%$ increase in the peak separation ($\sqrt{1+\mathcal{M}_{\rm II}^2}=1.49$ at $\mathcal{M_{\rm II}}\approx1.1$ for the fiducial run) which is not what we observe in the simulation. To understand correctly we need to note that core scattering of Ly$\alpha$ photons happens at small scale of the order of mean free path $\lambda_{\rm mfp}^{\rm core}=1/(\sigma_{\alpha,0}\nHI)$, or in the unit of the length of slab $\lambda_{\rm mfp}^{\rm core}/L=1/\tau_{\alpha,0}$. On the other hand, the rms Mach number refers to the turbulent velocity dispersion measured at the driving scale, i.e. at the length scale of slab $L$. The turbulence cascade transfers energy from large to small scale with the velocity dispersion decreasing to smaller scales. The velocity dispersion at the length scale $\ell$ is 
\begin{equation}
    \sigma_v(\ell)=c_{\rm s,II}\left(\frac{\ell}{\ell_s}\right)^{(n-1)/2},
\end{equation} 
where $\ell_s$ is the sonic length and $n$ is the power-law index of the velocity power spectrum $P_v(k)\propto k^{-n}$ ($n=5/3$ for subsonic and $n=2$ for supersonic turbulence, e.g. \citealt{Krumholz14} for a review). Therefore the velocity dispersion at the scale of core mean free path is
\begin{equation}
    \sigma_v(\lambda_{\rm mfp}^{\rm core})\sim c_{\rm s,II}\mathcal{M}_{\rm II}\tau_{\alpha,0}^{-(n-1)/2}.
\end{equation}
Thus, we argue that more appropriate inclusion of turbulence on the Doppler $b$-parameter is
\begin{equation}
    b=c_{\rm s,II}\sqrt{1+\mathcal{M}_{\rm II}^2\tau_{\alpha,0}^{-(n-1)}}\approx c_{\rm s,II},
\end{equation}
As $\tau_{\alpha,0}\sim10^4$ along the photoionized channels the turbulent velocity dispersion becomes negligibly small due to the turbulence cascade to the small scale of core scattering.\footnote{We have provided a physical argument. Clearly we do not resolve the full cascade down to the scale of mean free path in the simulation. Instead, it is limited to the grid scale of the simulation. However, the velocity dispersion becomes much smaller than the sound speed even after limiting the calculation only to the grid scale.} Thus, to the first order, the effect of turbulent on the peak separation remain small. A more precise estimate would defer from this as the turbulent velocity field is spatially correlated. We do not examine the precise influence of turbulence on spectral line \citep{Mihalas1978,Silantev06}, which may become increasingly important for a highly supersonic $\HII$ region ($\mathcal{M_{\rm II}}\gg1$).

\subsection{Shocked {H\,{\scriptsize \it I}} Shell and Fermi Acceleration?}

\begin{figure}
\includegraphics[width=\columnwidth]{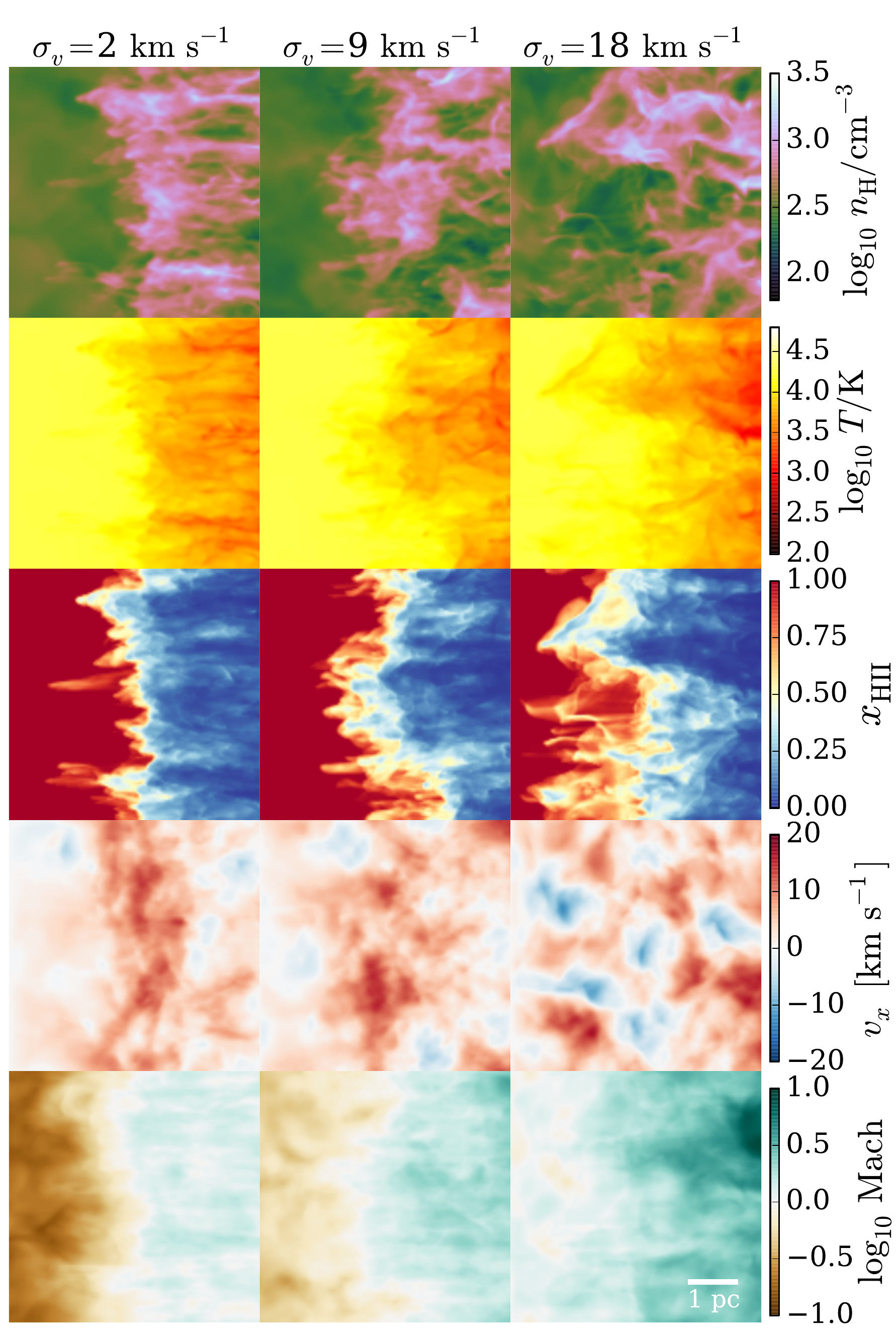}
\caption{The formation and destruction of the shocked $\HI$ shell in a turbulent medium. Projected maps of gas density $n_{\rm H}$, mass-weighted temperature $T$ and ionized fraction $\xHII$, $x$-velocity $v_x$, and the Mach number of the local rms velocity dispersion for the three simulations with different turbulent velocities (left: \texttt{V2S\_f2e11\_RHD}, middle: \texttt{V9S\_f2e11\_RHD}, right: \texttt{V18S\_f2e11\_RHD}) are shown. All the snapshots are at 0.10 Myr.}\label{fig:kinematics_map}
\end{figure}

We end by discussing the formation and destruction of the $\HI$ shell upon the propagation of the D-type I-front and the influence on the Ly$\alpha$ profile. In a homogeneous medium, the D-type I-front creates a shell of neutral hydrogen by the propagation of the photoheated shock front at the boundary of the $\HII$ region \citep{Whalen04,Hosokawa&Inutsuka06}, providing a probable mechanism for the classic shell model of Ly$\alpha$ line. However, in a turbulent medium, this $\HI$ shell is constantly perturbed by the turbulent flow as the I-front shock propagates the system, making its fate unclear.

Figure~\ref{fig:kinematics_map} shows the structure across the I-front for different turbulent velocities, indicating that a $\HI$ shell-like structure is disturbed for increasing turbulence. As the velocity of the shocked shell is of the order of sound speed $v_{\rm sh}\sim c_{s,\rm II}\simeq12.8T_4^{1/2}\rm~km~s^{-1}$, with increasing turbulent velocities, the timescale of the supersonic turbulent mixing becomes comparable to or faster than the speed that I-front shock sweeps up the neutral material $\sigma_v>v_{\rm sh}$. This causes the I-front shock to be constantly destroyed and mixed up with the neutral turbulent gas ahead of the I-front as soon as the $\HI$ shell develops. The simulations indicate this is the case; the $\HI$ shell is not formed for a higher Mach number. For a lower Mach number, the remnant of the shell structure is still visible and would survive of the order of Eddy turnover timescale.

This dispersal of the $\HI$ shell and the small shock velocity limit the efficiency of the blue wings and bumps production via Fermi-like acceleration of Ly$\alpha$ photons across shock fronts \citep{Neufeld&McKee88,Chung16}. In principle, the Fermi-accelerated Ly$\alpha$ contribute to blueshifted components at $\sim n v_{\rm sh}$ after $n$ crossings of a shock front \citep{Chung16}. However, as the shock velocity of the D-type I-front is small, even after several shock crossing the Fermi-accelerated Ly$\alpha$ photons only gain blueshifts of $\sim 25-80{\rm~km~s^{-1}}$. The gas outside the $\HII$ region is optically thick to these blueshifted photons with high column densities of $\NHI>10^{21}\rm~cm^{-2}$. Therefore the substantial frequency-space diffusion of several hundreds of $\rm~km~s^{-1}$ occurs by the subsequent scatterings through the ambient $\HI$ gas, effectively erasing the blueshifted component by the Fermi-acceleration. In fact, \citet{Neufeld&McKee88} estimated that for Fermi-accelerated blue peak to exceed the typical frequency diffusion of a slab, the shock velocity need to exceed $v_{\rm sh}>40T_{4}^{1/2}\rm~km~s^{-1}$. This value is hard to satisfy by the shock generated by the D-type I-front. Indeed, all our simulations show no noticeable effect of Fermi acceleration on the emergent Ly$\alpha$ profile. We expect that, in order for the production of Fermi-accelerated blue wings and bumps to be effective, other form of feedback such as supernova blastwave or stellar winds in the $\HII$ region is required to accelerate the shocked $\HI$ shell to a several hundreds of $\rm km~s^{-1}$ before being destroyed by turbulence.

\section{Discussion}\label{sec:discussion}

\subsection{Missing Physics: Dust}

We have employed an idealised RHD turbulence simulation in order to study the development of I-front and the associated LyC leakage and the Ly$\alpha$ line. We have ignored the effect of dust, metal line cooling, stellar wind, and gravity as well as the hydrodynamic instabilities associated with these missing physical processes. Here we discuss the limitations of our simulations and caveats.

Following the simple model assuming dust is perfectly mixed with gas, we find that the dust cross section per hydrogen atom is well approximated by
$\sigma_{\rm dust}(\lambda)\simeq5.3\times10^{-22}(\lambda/912$\,\AA$)^{-1}(Z/0.25\,Z_\odot)\rm~cm^2~H^{-1}$ over $800$\,\AA\,$<\lambda<1500$\,\AA~\citep{Gnedin08}.\footnote{This is based on a fit of \citet{Gnedin08} to \citet{Weingartner&Draine01} for Small Magellanic Cloud (SMC) type dust, assuming a linear dust-to-gas ratio $\propto Z$ normalized at the gas phase metallicity of the SMC, $Z=0.25~Z_\odot$ ($12+\log\rm O/H=8.1$, \citealt{Pagel03}).
Note that in reality a fraction of the dust is likely destroyed in the hot, ionized channels, and therefore this estimate overpredicts the impact of dust on the LyC leakage.}
The dust cross section is four order of magnitude lower than photoionization cross section. In photoionized channels, however, as the optical depth at Lyman limit is $\tau_L=\sigma_L\NHI\left[1+\sigma_{\rm dust}\NH/(\sigma_L\NHI)\right]$ where $\sigma_{\rm dust}\NH/(\sigma_L\NHI)=0.84(Z/0.25~Z_\odot)(\xHI/10^{-4})^{-1}$, the dust extinction can become comparable to the absorption by photoionization. For a system with a metallicity higher than $Z\geq0.3(\xHI/10^{-4})Z_\odot$, LyC leakage becomes increasingly suppressed by the absorption by dust. Strictly speaking, our idealised RHD turbulence simulation should therefore be only applicable for a metal-poor system with $Z<0.3(\xHI/10^{-4})Z_\odot$ where dust extinction becomes negligible. Furthermore, there are observational evidence that the dust-to-gas ratio may drop more rapidly than the linear extrapolation $\sigma_{\rm dust}\propto Z$ assumed here below $12+\log\rm O/H\leq8$ \citep{Remy-Ruyer14}. As we are concerned with metal-poor LyC-leaking systems ($12+\log\rm O/H\lesssim8$, i.e. $Z=0.2Z_\odot$, \citealt{Izotov2016,Izotov2018b}) and those analogous to reionisation-era galaxies \citep{Nakajima16,Senchyna17}, this condition is reasonably met. Therefore, our conclusion should be minimally affected by the complication by dust.

However, the effect of dust on the Ly$\alpha$ line properties can be complex. The abundance and distribution of dust controls the Ly$\alpha$ escape fraction \citep{Neufeld90}. Furthermore, dust can also affect the Ly$\alpha$ line shape as the level of dust attenuation is usually a function of the emergent frequency \citep{Neufeld90,Laursen09}. In the case when the dust is perfectly mixed with the neutral hydrogen, for instance, photons with an overall longer trajectory are more prone to destruction by dust. This can lead to an increased global asymmetry of the emergent spectrum, e.g. in the context of a homogeneous shell or slab the dust content can change the ratio of the peak fluxes for a double peaked profile \citep[e.g.][]{Gronke15}. For similar reasons, an increased dust content will make the individual peaks narrower (as photons further in the wing had a longer path length through neutral hydrogen and are thus more prone to destruction). This implies that the red peak asymmetry defined in \S~\ref{sec:peak_shape} will be lowered in such a scenario. However, while in Ly$\alpha$ radiative transfer studies it is frequently assumed that the dust number density is proportional to the neutral hydrogen, dust creation and destruction mechanisms are lead likely to a (Ly$\alpha$-affecting) dust distribution which is far more complex than this. For instance, dust may survive in ionized regions (given they are not too hot), or it clumps on small-scales due to streaming instabilities. We, therefore, leave the study of the imprint of dust on the Ly$\alpha$-LyC connection for future work.

\subsection{Missing Physics: \\ Metal Cooling, Winds, and Hydrodynamic Instabilities}

The metal line cooling introduces an interesting complication to the problem. If the neutral shocked shell ahead of the D-type I-front can radiatively cool by metal lines, the shocked gas is prone to fragmentation via a so-called {\it thin-shell instability} \citep[][and refrences therein]{Whalen&Norman11}. This facilitates to open cranks in the shell and allows further escape of radiation through the channels. The growth of the thin-shell instability depends the availability of efficient cooling mechanisms beyond primordial hydrogen and helium coolings \citep{Whalen&Norman2008b,Whalen&Norman2008a}. Since we have ignored the effect of cooling by metal lines, such mode of instability is inhibited by design.

There are other instabilities and winds that may influence the structure of the $\HII$ region. Under a gravitational field, the Rayleigh-Taylor instability can also develop as rarefied ionized gas pushed the dense ambient neutral medium \citep[e.g.][]{Jacquet&Krumholz11,Jiang13,Park14}, which, if operate on the relavent timescale, can further contribute to the fluctuation of the I-fronts. The metal-line driven stellar winds of massive stars can inject fast outflow into the $\HII$ region, which can produce a hot X-ray emitting gas of $\sim10^6\rm~K$ by the clumping and shocks in the outflowing gas amplified by the line-deshadowing instability \citep[e.g.][]{Owocki2015}. The wind-blown bubble can further evacuate the gas through openings of low column density channels, in which both hot gas and radiation can leak out more efficiently \citep{Rogers&Pittard13}.  Again this will likely amplify the fluctuation of the I-fronts and the multiphase structure of the $\HII$ regions that are initially seeded by turbulence. Inclusion of metal line cooling, stellar winds, and gravity would therefore lead to the amplification of the I-front inhomogeneities, probably allowing more escape of radiation, and alter the emergent Ly$\alpha$ spectra. However, as we have studied the the development of I-front in a {\it driven} turbulence medium, the growth of the instabilities is subject to constant destruction by the external turbulent mixing over the Eddy turnover timescale. Thus, it still remains unclear what the exact impacts of the instabilities and winds under an external source of turbulence are and how they ultimately influence the LyC leakage and Ly$\alpha$ spectra.

\subsection{Influence of Turbulence on Other Nebular Lines}

The intense nebular $[\OIII]$+H$\beta$ and $\HeII$ emission lines are often associated characteristics of reionization-era galaxies and LyC-leaking galaxies. As discussed in Section \ref{sec:result}, supersonic turbulence in the $\HII$ region can induce density fluctuations. Since the nebular recombination lines scale as $\propto n_e^2$, this may lead to the enhanced nebular emission line luminosity,
\begin{align}
    L_{\rm neb}&=\gamma_{\rm neb}(T)\bar{n}_e^2 \mathcal{C}(\mathcal{M}_{\rm II})V_{\rm HII}, \nonumber \\
    &\approx\gamma_{\rm neb}(T)\bar{n}_e^2\left[1+\left(\frac{\mathcal{M}_{\rm II}}{3}\right)^2\right]V_{\rm HII},
\end{align}
where $\gamma_{\rm neb}(T)$ is the emission coefficient and $V_{\rm HII}$ is the volume of the $\HII$ region. For example, a high velocity dispersion medium with $\mathcal{M_{\rm II}}=3$ (i.e. $\sigma_v\approx38\rm~km~s^{-1}$) could lead to a factor of two boost in the emission line luminosity. The equivalent width of the emission line could also be boosted accordingly.

\citet{Gray&Scannapieco17} have examined the impact of turbulence on the line ratios in detail by incorporating the non-equilibrium chemistry. They have shown that a high turbulent velocity can generally increase the nebular emission line ratios due to the the associated temperature fluctuations, which mimics the effect of harder stellar sources in the locus of nebular diagnostic diagrams. Because we have neglected metals and assumed the isothermal equation of state, this effect is missing from our simulations. These clumping and thermal fluctuations may complicate the relation between LyC escape fractions and $[\OIII]/[\OII]$ ratios \citep{Izotov2016,Faisst16} and may contribute to the observed scatter in the $f_{\rm esc}^{\rm LyC}-[\OIII]/[\OII]$ correlation \citep{Naidu18,Bassett19}.

For Ly$\alpha$ line profiles of Green peas galaxies, when interpreted with homogeneous shell models, the required intrinsic Ly$\alpha$ line width exceed that of observed H$\beta$ line width ($\sim130-230\rm~km~s^{-1}$) for successful fit to the data, which causes problematic fits when the consistency with H$\alpha$, H$\beta$, and/or $[\OIII]~\lambda5007$ line widths is required (\citealt{Yang16,Yang17,Orlitova18}, but see \citealt{Gronke18}). For a turbulent $\HII$ region, a narrow Ly$\alpha$ injection at line center can produce both narrow peak separation through photoionized channels and broad wing component by multiple scatterings through optically thick channels. As the turbulence line broadening of the ionized gas is of the order of tens of $\rm km~s^{-1}$, the observed H$\beta$ line width can still accommodate the turbulence broadening within individual $\HII$ regions as well as the contributions from thermal broadening and the velocity dispersion of multiple $\HII$ regions in a galaxy.

Overall, it is important to include the effect of turbulent $\HII$ regions on the nebular emission lines in stellar+$\HII$ region population synthesis modelling to understand the observed relations between LyC, Ly$\alpha$, and nebular emission line properties. 

\subsection{Scale of LyC Leakage: Observational Test with the Magellanic Systems and Local Blue Compact Dwarfs}

The picture that LyC leakage from the ISM of a galaxy is controlled by the escape fractions through molecular clouds assumes that a major source of opacity in the ISM comes from GMCs rather than diffuse gas in between them, arguing that $f_{\rm esc,gal}^{\rm LyC}\approx \langle f_{\rm esc,GMC}^{\rm LyC}\rangle$ where
\begin{equation}
\langle f_{\rm esc,GMC}^{\rm LyC}\rangle\equiv\frac{\displaystyle\int \dot{N}_{\rm ion}(M_{\rm cl})f^{\rm LyC}_{\rm esc}(M_{\rm cl})\frac{d\mathcal{N}}{dM_{\rm cl}}dM_{\rm cl} }{\displaystyle\int \dot{N}_{\rm ion}(M_{\rm cl})\frac{d\mathcal{N}}{dM_{\rm cl}}dM_{\rm cl}},
\end{equation}
where $f_{\rm esc,gal}^{\rm LyC}$ is the galactic escape fraction that is averaged over an entire galaxy, $f^{\rm LyC}_{\rm esc}(M_{\rm cl})$ is the escape fraction from a GMC of mass $M_{\rm cl}$, and $d\mathcal{N}/dM_{\rm cl}$ is the mass distribution of the GMCs. We discuss a way to test the spatial scale responsible for LyC leakage and the associated feedback in the $\HII$ regions. 

The galactic escape fraction can be inferred from the diffuse H$\alpha$ emission from the CGM of a galaxy \citep{Bland-Hawthorn&Maloney99,Mas-Ribas17}. This approach was applied to the two local dwarf galaxies, Small and Large Magellanic Clouds, of the Milky Way by \citet{Barger13}. They find that the diffuse H$\alpha$ emission from the Magellanic Bridge -- the diffuse gas inbetween the Magellanic Clouds -- shows an excess H$\alpha$ surface brightness from the diffuse gas that cannot be explained by the photoinization from the escaping ionizing radiation from the the Milky Way (cf. a few per cent $f_{\rm esc,gal}^{\rm LyC}$, \citealt{Bland-Hawthorn&Maloney99,Bland-Hawthorn&Maloney01}) and the extragalactic UV background \citep{Haardt&Madau01}. By attributing the H$\alpha$ emission by the photonization due to the LyC photons leaking from the Magellanic Clouds, they placed an upper limits to the escape fraction of $f_{\rm esc,gal}^{\rm LyC}<4.0\%$ for LMC and $f_{\rm esc,gal}^{\rm LyC}<5.5\%$ for SMC. As this measures the LyC photons arrived at the Magellanic Bridge after escaping out of the Magellanic Clouds, this provides a measure of `galactic' escape fraction. 

On the other hand, escape fractions from individual $\HII$ regions in a galaxy can be measured by estimating the direct ionizing photon production rate of the massive stars (by direct stellar spectroscopy or spectral energy distribution fitting) and the recombination rate in each $\HII$ region. As the nebular H$\alpha$ luminosity of the $\HII$ region is proportional to the amount of ionising photons absorbed (recombined) in the region, each escape fraction can be estimated by $f_{\rm esc,GMC}^{\rm LyC}=(\dot{N}_{\rm ion}-\dot{N}_{\rm rec})/\dot{N}_{\rm ion}$. The application of the method to the LMC indicates that the brightest $\HII$ region, 30 Doradus, has an escape fraction of $f_{\rm esc,GMC}^{\rm LyC}\sim6^{+55}_{-6}\%$ \citep{Doran13}. The individual $\HII$ region's escape fractions varies enormously from an object to an object; for example, the $\HII$ region complexes N44 and N180 show escape fractions as large as $f_{\rm esc,GMC}^{\rm LyC}\sim40-80\%$ \citep{McLeod18}. Each $\HII$ region can be classified via line ratios to ionization- or density-bound nebula using the ionization parameter mapping and, when the indirect measure of $f_{\rm esc,GMC}^{\rm LyC}$ is averaged over the all $\HII$ regions, it has been suggested that the population-averaged escape fraction is $\langle f_{\rm esc,GMC}^{\rm LyC}\rangle\sim42\%$ for the LMC \citep{Pellegrini12}. 

If this value of $\langle f_{\rm esc,GMC}^{\rm LyC}\rangle$ is compared to the estimate of $f_{\rm esc,gal}^{\rm LyC}$, at face value, it seems that additional $\sim90\%$ of LyC absorption by the diffuse ISM between the $\HII$ regions is required to give the observed galactic escape fraction. Unfortunately, the uncertainties including the recently revised extragalactic UV background value \citep[e.g.][]{Shull15,Khaire19} and various differing assumptions make it difficult to draw a definitive conclusion. The above argument nonetheless should illustrate a way in which the mechanism of LyC leakage could be tested observationally. Given the similarity of the central star-forming region NGC2070 of 30 Doradus to the local LyC-leaking and Green Pea galaxies in their emission line and star formation properties \citep{Crowther17}, the modern integral-field spectroscopic census of $\HII$ regions of the Magellanic Clouds and the revised homogeneous analysis of $f_{\rm esc,GMC}^{\rm LyC}$ and $f_{\rm esc,gal}^{\rm LyC}$ will be extremely useful to examine the physical mechanism and scale of LyC leakage, with which the system's LyC leakage can also be correlated with the stellar feedback mechanisms in the $\HII$ regions including photoionization heating, radiation pressure, and stellar winds by massive stars \citep{Lopez11,Lopez14,McLeod18}. 

The similar method should be applicable for nearby blue compact dwarf galaxies for which the individual $\HII$ regions and the diffuse H$\alpha$ emission from the halos may be examined by narrow band imaging and/or deep integral field spectroscopy. The closely related approach was already taken by \citet{Weilbacher18} who examined the LyC leakage from the $\HII$ regions in the Antennae galaxy and \citet{Menacho19} who reported the diffuse H$\alpha$ halo around a LyC-leaking galaxy, Haro 11. Such observational sample should provide a valuable spatially resolved reference sample for reionization-era galaxies to test the role of turbulence and stellar feedback on LyC leakage and to help the interpretation of future observations with {\it JWST} and Extremely Large Telescopes (ELT).

\section{Conclusions}\label{sec:conclusion}

We have examined the physical origin of LyC leakage and the associated Ly$\alpha$ spectra through turbulent $\HII$ regions using fully-coupled radiation hydrodynamic simulations, representing the growth of the ionization front in a GMC. Using a series of RHD turbulence simulations with \textsc{ramses-rt} in a plane-parallel geometry where the turbulence is constantly driven on the large scale ($\sim$ parsec), we have computed LyC escape fractions and calculated the associated Ly$\alpha$ spectra using the Monte-Carlo radiative transfer code \textsc{tlac}, whereby their correlations with $\HI$ covering fraction, gas kinematics, and spectral hardness of ionising sources and the roles of turbulence and radiative feedback are examined in detail.

We find that LyC photons escape through turbulence-generated low column density channels in a $\HII$ region which are evacuated efficiently by radiative feedback induced by shocks due to the photoionization heating across the D-type I-fronts. Both turbulence {\it and} radative feedback are key ingredients for regulating the LyC leakage. Because both processes can operate just after the birth of massive stars, high LyC leakage can be achieved at early times before the onset of supernova feedback. This mechanism generates a time variable LyC escape fraction which anti-correlates with the $\HI$ covering fraction, and correlates with turbulence velocities and spectral hardness of the sources. As the LyC photons recombine through the low column density channels, the resulting escape fraction deviates from the $1-f_{\rm cov}$ expectation. This confirms that while a low $\HI$ covering fraction is a necessary condition for high LyC leakage, it only provides an upper limit to the actual escape fraction.  The turbulent gas kinematics influences the escape fraction by modifying the densities through the photoionized channels, which generally lead to increasing escape fractions with higher turbulence velocities at a given $\HI$ covering fraction and spectral hardness.  

The emergent Ly$\alpha$ spectra correlates with the LyC leakage mechanism, reflecting the porosity and multiphase structure of the turbulent $\HII$ regions. Ly$\alpha$ photons funnel through the photoionized channels in which LyC photons escape. Depending on the availability of the ionization- and density-bound channels which are regulated by turbulence and radiative feedback, the $\HII$ regions produce diverse Ly$\alpha$ spectral morphology including narrow double-peaked profiles. For a LyC-leaking $\HII$ region, instead of the total $\HI$ column density of the system, the Ly$\alpha$ peak separation is set by the residual $\HI$ column density and temperature of the photoionized channels. This means that it is possible to have a system with a narrow Ly$\alpha$ peak separation and high LyC leakage, while retains a relatively high $\HI$ mass on average.  The peak asymmetry reflects the porosity of the $\HII$ region. A low asymmetry is often associated with both density- and ionization-bound dominated systems whereas a high asymmetry is associated with a mixed system of the two phases as the multiple routes of Ly$\alpha$ escape are available. It may therefore be possible to distinguish anisotropic LyC leakage through holes and isotropic leakage from a fully density-bound medium using the red peak asymmetry as a diagnostic. 

In summary, radiative transfer through LyC-leaking $\HII$ regions in turbulent molecular clouds provides an appealing picture to interpret the observed Ly$\alpha$ spectra of LyC-leaking galaxies and provide a natural mechanism to explain some of the observed Ly$\alpha$ spectral characteristics. This provides an appealing hypothesis to explain high LyC leakage and Ly$\alpha$ spectra observed in very young star-forming galaxies in the local Universe without need of extreme galactic outflows or supernova feedback. Although the diffuse ISM and CGM will clearly add additional complexities to the observed LyC leakage and Ly$\alpha$ spectral properties, it is worth emphasizing the importance of the physical processes in $\HII$ regions and GMCs that are poorly resolved components in galaxy formation simulations and often treated only in a simplified manner in stellar population synthesis tools and photoionization modeling used for the analysis of observed galaxies.

The connection between turbulence and stellar feedback in $\HII$ regions, LyC leakage, Ly$\alpha$ spectra, and nebular emission lines is testable with integral field spectroscopic studies of blue compact dwarf galaxies and the $\HII$ regions in the Magellanic Clouds. These targets are valuable laboratories for reionization-era systems. In order to correctly interpret the upcoming {\it JWST} and ELT observation of high-redshift galaxies, it is critical to incorporate the impact of turbulent $\HII$ regions and the associated LyC, Ly$\alpha$ and rest-frame UV-to-optical line properties self-consistently in the stellar population synthesis modeling. Further theoretical and observational investigations are needed. The future prospects include providing a spectral library of $\HII$ regions using RHD simulations for population synthesis and the calibration against the spatially resolved studies of $\HII$ regions and nearby dwarfs as analogs of reionization-era galaxies.

\acknowledgments

We thank Sam Geen, Brant Robertson, Jeremy Blaizot, and Richard Ellis for helpful discussions and comments. Special thanks goes to Joki Rosdahl for discussion and technical questions regarding \textsc{ramses-rt} as well as Romain Teyssier and the \textsc{ramses} user community for making the code public and actively maintaining it. KK acknowledge financial support from European Research Council Advanced Grant FP7/669253. MG was supported by by NASA through the NASA Hubble Fellowship grant \#HST-HF2-51409 awarded by the Space Telescope Science Institute, which is operated by the Association of Universities for Research in Astronomy, Inc., for NASA, under contract NAS5-26555. MG thanks the Institute for Theoretical Astrophysics in Oslo for their hospitality. This work is based on observations made with the NASA/ESA Hubble Space Telescope, obtained from the data archive at the Space Telescope Science Institute. The simulation was undertaken using the UCL Grace High Performance Computing Facility (Grace@UCL) and we thank the associated support services. 

\software{RAMSES-RT \citep{Teyssier02,Rosdahl13}, TLAC \citep{Gronke14}, Pynbody \citep{pynbody}, Astropy \citep{AstropyI,AstropyII}}

\section*{Appendix A \\ Turbulence forcing method}\label{app:A}
We implemented the turbulence forcing scheme to \textsc{ramses-rt} to enable the fully coupled radiation hydrodynamical simulations of a driven turbulence medium. In our implementation we perturb the momentum and the gas energy density ten times per eddy turnover timescale $T=L/(2V_{\rm rms})$ where $V_{\rm rms}$ is a simulation parameter. Every updates are given by
\begin{align}
(\rho\boldsymbol{v})^{n+1}=(\rho\boldsymbol{v})^n+\rho^n\delta\boldsymbol{v} \\
E^{n+1}=E^n+(\rho\boldsymbol{v})^n\cdot\delta\boldsymbol{v}
\end{align}
where $\delta\boldsymbol{v}$ is a Gaussian random field. We have chosen this reduced frequency of turbulence forcing update scheme so as to reduce computational cost. We generate the velocity perturbation field using two different methods.

The first method is the one used by \citet{Robertson&Goldreich12,Robertson&Goldreich18}, which is in turn based on \citet{Kritsuk07}. The random velocity perturbation field is generated by
\begin{equation}
\delta\tilde{\boldsymbol{v}}(\boldsymbol{k})=\hat{\sigma}_v(\boldsymbol{k})\boldsymbol{\mathsf{P}}_\zeta(\boldsymbol{k})\boldsymbol{n}(\boldsymbol{k})
\end{equation}
where $\boldsymbol{n}(\boldsymbol{k})$ is the Fourier transform of a white noise field, $\hat{\sigma}_v(\boldsymbol{k})$ is the injection power spectrum, which we chose a flat spectrum over $1<k<k_0$ but otherwise zero. The normalisation of the injection spectrum is chosen such that the rms of the forcing field $\langle\left|\delta\boldsymbol{v}\right|^2\rangle^{1/2}=V_{\rm rms}$. A Helmholz decomposition is done by applying the projection tensor $\boldsymbol{\mathsf{P}}$, for which each component is given by \citep{Federrath10}
\begin{equation}
\mathsf{P}_{ij}(\boldsymbol{k})=\zeta\delta_{ij}+(1-2\zeta)\frac{k_i k_j}{|k|^2},
\end{equation}
where $\delta_{ij}$ is the Kronecker delta. The white noise field is newly generated at every update times. Thus, this method generates forcing fields that are completely independent in time. Every turbulent perturbations are indepedent from previous timesteps. The resulting velocity dispersion of the turbulent flow is then measured directly from the simulation output.

\begin{figure}
\hspace{0.5cm}
\includegraphics[width=\columnwidth]{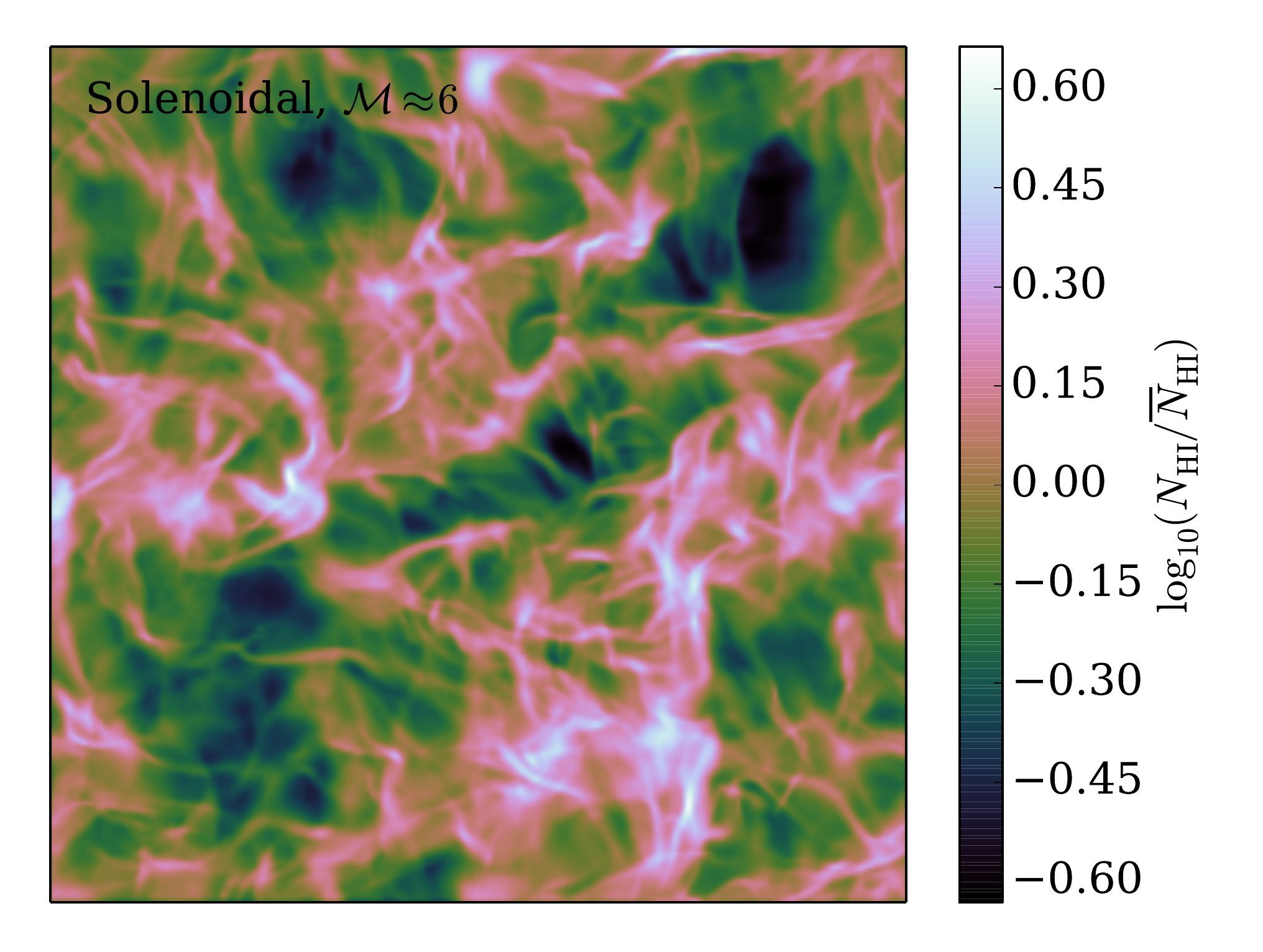}
\caption{The map of $\HI$ column density fluctuations in a driven isothermal turbulence simulation of $\mathcal{M}\approx6$ flow. The $256^3$ simulation with solenoidal forcing and the box size is 5 pc on a side.}\label{fig:test1}
\end{figure}
\begin{figure}
\hspace{-0.3cm}
\includegraphics[width=\columnwidth]{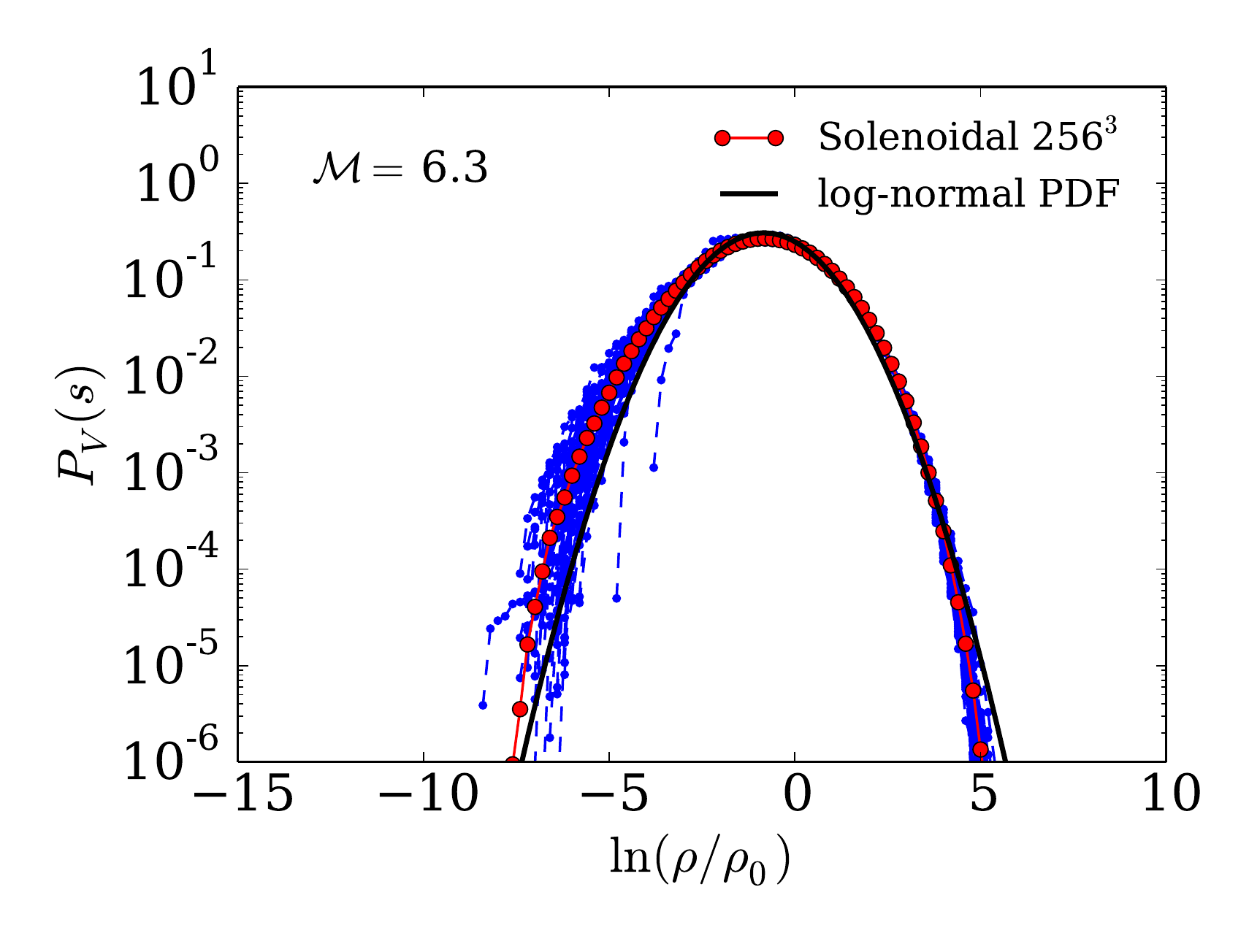}
\vspace{-0.5cm}
\caption{The volume-weighted PDF of gas density contrasts. The average PDF after $>1T$ is indicated by the red line, and the PDF of each snapshot is shown in blue. The log-normal PDF indicated by the black line.}\label{fig:test2}
\end{figure}

To test the forcing algorithm, we run $256^3$ uniform grid hydrodynamical simulation of supersonic isothermal turbulence in a periodic box of size 5 pc on a side. The simulation was run for five turnover time $5T$ and the outputs are recored every $0.1T$ intervals. Figure \ref{fig:test1} shows the map of projected column densities, which visually in agreement with the known morphology of turbulent density fluctuations \citep[e.g.][]{Federrath10}. For a more quantitative test, the volume-weighted density distribution function (PDF) $P_V(s)$ averaged over all snapshots at $t>1T$ (red line) is shown in Figure \ref{fig:test2}. The log-normal PDF fits very well to the simulated distribution. Although some deviations are found at the both low- and high-density tails of the distribution, such deviations are known in previous studies. The departure from the log-normal PDF is likely due to the limited spatial resolution at the high-density tail and the intermittency at the low-density tail, which increases for higher Mach numbers.

Overall, our forcing algorithm in \textsc{ramses-rt}  agrees with known results of turbulence properties. We have repeated the test with $128^3$ grid resolution and found an almost identical result. For the RHD turbulence simulation, we therefore adopt the $128^3$ resolution throughout the paper.

\section*{Appendix B \\ HST/COS sample}\label{app:COS}

\begin{table}
\centering
\caption{{\it HST}/COS LyC-leaking sample at $z\sim0.3$}\label{table:cos}
\begin{tabular}{llll}
\hline\hline
Name & $f_{\rm esc}^{\rm LyC}$ & $\Delta v_{\rm peak}$ & $A_f$  \\
     & [$\%$] & [$\rm km~s^{-1}$] & [-] \\
\hline
J0901+2119  & $2.7\pm0.7^a$  & $345.0\pm12.5^a$  & $2.09\pm0.19$ \\
J1011+1947  & $11.4\pm1.8^a$ & $276.4\pm5.4^a$ & $2.31\pm0.30$ \\
J1243+4646  & $72.6\pm9.7^a$ & $143.4\pm4.0^a$ & $3.31\pm0.45$ \\
J1248+4259  & $2.2\pm0.7^a$  & $283.8\pm15.9^a$ & $4.27\pm0.65$ \\
J1256+4509  & $38.0\pm5.7^a$ & $239.4\pm10.5^a$ & $1.61\pm0.24$	\\
J1154+2443	& $46.0\pm2.0^b$ & $199.0\pm10.0^{b,\!\ast}$ & $2.79\pm0.31$ \\
J0925+1403	& $7.20\pm0.8^c$ & $310.0\pm10.0^{d,\!\ast}$ & $3.44\pm0.46$ \\
J1152+3400	& $13.2\pm1.1^c$ & $270.0\pm10.0^{d,\!\ast}$ & $4.01\pm0.54$ \\
J1333+6246	& $5.60\pm1.5^c$ & $390.0\pm10.0^{d,\!\ast}$ & $1.51\pm0.20$ \\
J1442-0209	& $7.40\pm1.0^c$ & $310.0\pm10.0^{d,\!\ast}$ & $2.28\pm0.10$	\\
J1503+3644	& $5.80\pm0.6^c$ & $430.0\pm10.0^{d,\!\ast}$ & $2.36\pm0.34$ \\
\hline
\multicolumn{4}{l}{
$^a$ \citet{Izotov2018b}
$^b$ \citet{Izotov2018a}} \\
\multicolumn{4}{l}{
$^c$ \citet{Izotov2016}
$^d$ \citet{Verhamme17}} \\
\multicolumn{4}{m{8cm}}{$^\ast$ When the error in the peak separation is not explicitly stated, the $\pm10\rm~km~s^{-1}$ uncertainty corresponding to the COS spectral resolution $R=15000$ is used.}
\end{tabular}
\end{table}

In order to compare the simulation with the {\it HST}/COS observation of low redshift LyC-leaking galaxies, we have retrieved the reduced COS G160M spectra of $z\sim0.3$ \citet{Izotov2016,Izotov2018a,Izotov2018b} sample from the MAST archive (GO 14635: Izotov, GO 13744: Thuan). For the measurement of LyC escape fractions and Ly$\alpha$ peak separations we have used the reported values from \citet{Izotov2016,Izotov2018a,Izotov2018b} and \citet{Verhamme17}, which are tabulated in Table~\ref{table:cos}. For the red peak asymmetry parameter $A_f$, we have measured the quantity directly from the archival COS spectra after binning to the resolution matched to $R=15000$. We first identified the wavelengths of the red and blue peaks, $\lambda_{\rm peak}^{\rm red}$ and $\lambda_{\rm peak}^{\rm blue}$, from the maximum of each component. The valley is located as the minimum between the two peaks, corresponding to wavelength $\lambda_{\rm valley}$. The red peak asymmetry parameter is then computed as the ratio of right-to-left flux of the red peak, $A_f=\left.\int_{\lambda^{\rm red}_{\rm peak}}^{\lambda_{\rm max}} f_\lambda d\lambda\right/\int_{\lambda_{\rm valley}}^{\lambda^{\rm red}_{\rm peak}} f_\lambda d\lambda$, where $\lambda_{\rm max}$ is the maximum wavelength for the red peak which is chosen to be $\lambda_{\rm max}=1220$~\AA. The asymmetry parameters for all the objects are shown in Table~\ref{table:cos}. We use these tabulated values for our comparison with the simulation.

\bibliography{references}

\end{document}